\documentclass[twocolumn,showpacs,preprintnumbers,amsmath,amssymb,prb]{revtex4}

\usepackage{graphicx}
\usepackage{dcolumn}
\usepackage{bm}
\usepackage{longtable}

\begin{document}

\title{Spectral moment sum rules for strongly correlated electrons
in time-dependent electric fields}

\author{V.~M~Turkowski} \email{turk@physics.georgetown.edu}
\author{J.~K.~Freericks}\email{jkf@physics.georgetown.edu} 
 \affiliation{Department of Physics, Georgetown University,
Washington, D.C. 20057 USA}

\date{\today}

\begin{abstract}
We derive exact operator average expressions for the first two spectral moments
of nonequilibrium Green's functions 
for the Falicov-Kimball model and the Hubbard model
in the presence of a spatially uniform,
time-dependent electric field. The moments are similar to the well-known
moments in equilibrium, but we extend those results to systems in arbitrary
time-dependent electric fields.
Moment sum rules can be employed to estimate the accuracy of 
numerical calculations; we compare our theoretical results to numerical
calculations for the nonequilibrium dynamical mean-field theory solution of the
Falicov-Kimball model at half-filling. 

\end{abstract}

\pacs{71.27.+a, 71.10.Fd, 71.45.Gm, 72.20.Ht}


\maketitle

\section{\label{sec:level1} Introduction}

The problem of strong electron correlation is one of the most
challenging problems in condensed matter physics. It is interesting because
many materials with important properties for applications derive those
properties from the delicate balance between minimizing the kinetic and the
potential energies in strongly correlated materials. Most theoretical work on
this problem has focused on equilibrium properties and linear response, with 
only limited work available on the nonequilibrium system (which is most easily
attained when driven by an external electric field).
Two commonly studied models of strong electron correlation are the Hubbard 
model\cite{Hubbard}
and the Falicov-Kimball model\cite{FalicovKimball}  
(equivalent to the Hubbard model
with zero hopping parameter for the down spin electrons).
Despite tremendous efforts, the exact
equilibrium solution of these models is known only in some 
limiting cases like one dimension, where
the Bethe anzatz technique\cite{LiebWu} can be successfully applied to the
Hubbard model,
and in the infinite-dimensional case where both models can be 
solved\cite{Georges,Freericks} with dynamical mean-field theory (DMFT). 

Since the exact solution of these problems is challenging to attain, exact
results in the form of sum rules can be quite valuable in determining the
fidelity of different approximation techniques (be they analytic, variational,
perturbative, or numerical approximations). 
This problem was analyzed in equilibrium for the Hubbard model by
Steve White\cite{white} and used to check
the accuracy of a quantum Monte Carlo solution to the two-dimensional Hubbard 
model.

Can similar results be found for nonequilibrium situations, like the case
of a strongly correlated material in an
uniform external electric field (with arbitrary time dependence)? The answer
to this question has, surprisingly, not been discussed much in the literature.
It is well known that the canonical anticommutation relation for fermion
creation and annihilation operators leads to the integral of the spectral
function being 1 in both equilibrium and nonequilibrium situations.  It is 
also well known that the conventional proof that the spectral function
is nonnegative in equilibrium (arising from a Lehmann representation),
does not apply to the nonequilibrium case, so the spectral function can
be negative.  It is also known that in the limit of a steady state, the
spectral function recovers its nonnegativity.  In this work, we examine 
the theoretical problem of the first few moments of the spectral functions
in the presence of an external electric field.  It turns out that the
equilibrium results for the first few moments are quite similar to the
nonequilibrium results, implying they can be used effectively to determine
the accuracy of different approximation techniques in solving nonequilibrium
problems.

Dynamical mean-field theory has been employed to solve many of the 
models of strongly correlated electrons.  To date, most of this work
has focused on equilibrium properties.  Schmidt and Monien\cite{SchmidtMonien},
made a first attempt to solve nonequilibrium DMFT via second-order perturbation
theory for the Hubbard model.  Their theoretical formulation evaluated the
case of a spatially uniform, but time dependent scalar potential, which 
unfortunately does
not correspond to any electric field. More recently, a generalization
of the Brandt-Urbanek solution\cite{brandt_urbanek} for the localized
electron spectral function allows for an exact numerical solution of the
nonequilibrium problem for the Falicov-Kimball model\cite{Nashville} within
the DMFT framework.  The
procedure works directly in time by discretizing the continuous matrix operators
and solving the nonlinear equations by iterating matrix operations on the
discretized operators. The approach has been tested against the equilibrium
solution\cite{Nashville}, and nonequilibrium results for the quenching of
Bloch oscillations will be presented elsewhere.

In this contribution, we derive operator identities for the first two spectral
moments of the nonequilibrium Green's functions when the strongly correlated
material is in the presence of a spatially uniform electric field with
arbitrary time dependence. These identities are found for both the 
Falicov-Kimball and the Hubbard model.
Our results are valid for any spatial dimensionality. In the general case,
the moments depend on time, which is expected because the field can
be turned on at any time; surprisingly, the first two moments of the local 
retarded Green's function are time independent, and have the same form as
in equilibrium. This last result is particularly
surprising because the local retarded
Green's function is a nontrivial  oscillating function of time in the 
noninteracting case\cite{Turkowski}.

The rest of the paper is organized as follows. We derive analytical
expressions for the spectral moments for the Falicov-Kimball model
and the Hubbard model in Section II.  
The application of nonequilibrium DMFT to the Falicov-Kimball model
follows in Section III.  Section IV contains results for the Green's functions 
and spectral moments in this case.
A summary of our results and our conclusions appear in Section V.

\section{\label{sec:level2} Formalism for the spectral moments}

We begin our formal developments with a derivation of exact expressions
for the zeroth, first and second spectral moments
of the retarded and lesser Green's functions;
the analysis is performed 
for the spinless Falicov-Kimball model and the spin one-half Hubbard model
in an external arbitrary time-dependent spatially uniform electric field.
The Hamiltonian for both models has the following form (in the absense of
an external field):
\begin{eqnarray}
{\mathcal H}(0)&=&-\sum_{ij}t_{ij}c_{i}^{\dagger}c_{j}
-\sum_{ij}t_{ij}^{f}f_{i}^{\dagger}f_{j}
-\mu\sum_{i}c_{i}^{\dagger}c_{i}\nonumber\\
&-&\mu_{f}\sum_{i}f_{i}^{\dagger}f_{i}
+U\sum_{i}f_{i}^{\dagger}f_{i}c_{i}^{\dagger}c_{i}.
\label{H}
\end{eqnarray}
In the case of the Falicov-Kimball model,\cite{FalicovKimball}~the 
Hamiltonian in Eq.~(\ref{H})
describes a system which consists of two kinds of spinless electrons:
itinerant $c$-electrons with a nearest neighbor hopping
matrix $t_{ij}$ and localized $f$-electrons with a hopping
matrix equal to zero ($t_{ij}^{f}=0$). We normally take the hopping matrix to 
be between nearest neighbors only, but this is not a requirement.  We do
assume the matrix elements are all real and that the hopping matrices are
Hermitian. The on-site interaction between the two electrons is equal to $U$.
Double occupation by a $c$ or $f$ electron is forbidden by the Pauli
exclusion principle.  The chemical potentials are $\mu$ and $\mu_{f}$ for the
$c$- and $f$-electrons, respectively.
We will set $\mu_{f}=0$ in our calculations; it plays no role in the 
spectral moments of the $c$ particles, which we will be evaluating
in this contribution. The spectral moments of the $f$-electrons in
equilibrium were worked out in Ref.~\onlinecite{f_spectral}.
In the case of the Hubbard model,\cite{Hubbard}~the Hamiltonian in Eq.~(\ref{H})
describes a system of spin-up $c\equiv c_{\uparrow}$-electrons, 
and spin-down $f\equiv c_{\downarrow}$-electrons
with equal hopping matrix elements $t_{ij}=t_{ij}^{f}$
and chemical potentials $\mu =\mu_{f}$.
The local Coulomb repulsion between electrons with different
spins is $U$.

The electric field ${\bf E}({\bf r}, t)$ can be described by a vector potential
${\bf A}({\bf r}, t)$ in the Hamiltonian or temporal
gauge (where the scalar potential vanishes):
\begin{equation}
{\bf E}({\bf r}, t)=-\frac{1}{c}\frac{\partial {\bf A}({\bf r}, t)}
{\partial t}
\label{Electricfield}
\end{equation}
We assume that vector potential ${\bf A}({\bf r}, t)$
is smooth enough, that the magnetic field produced by ${\bf A}({\bf r}, t)$
can be neglected. For simplicity we shall assume that
the vector potential is spatially uniform (independent of {\bf r}).
Note that if we have any time dependence to the electric field, then
neglecting the magnetic field will violate Maxwell's equations.  In most
situations these magnetic field effects are small enough that they can
be neglected in a first analysis, and added back later via either perturbative
or gradient-based approaches.
The electric field is coupled to the electrons via the Peierls 
substitution,\cite{Jauho}~which 
involves modifying the hopping matrix elements by a phase that depends on 
the line integral of the vector potential:
\begin{eqnarray}
t_{ij}&\rightarrow& t_{ij}\exp\left[
-\frac{ie}{\hbar c}\int_{{\bf R}_{i}}^{{\bf R}_{j}}{\bf A}({\bf r}, t)d{\bf r}
\right] \nonumber\\
&=&t_{ij}\exp\left[
-\frac{ie {\bf A}(t)}{\hbar c}
\cdot ({\bf R}_{i}-{\bf R}_{j})
\right] ,
\label{Peierls1}
\end{eqnarray}
\begin{eqnarray}
t_{ij}^{f}&\rightarrow& t_{ij}^{f}\exp\left[
-\frac{ie}{\hbar c}
\int_{{\bf R}_{i}}^{{\bf R}_{j}}{\bf A}({\bf r}, t)d{\bf r}
\right] \nonumber\\
&=&t_{ij}^{f}\exp\left[
-\frac{ie {\bf A}(t)}{\hbar c}
\cdot ({\bf R}_{i}-{\bf R}_{j})
\right] .
\label{Peierls2}
\end{eqnarray}
The second line in each equation follows for spatially uniform vector potentials.
Note that the Hamiltonian in a field, $\mathcal{H}({\bf A})$, is identical in
form to that shown in Eq.~(\ref{H}), but it uses the hopping matrices in
Eqs.~(\ref{Peierls1}) and (\ref{Peierls2}).  Note also that $t^f=0$ for the
Falicov-Kimball model.

The ``Peierls substituted''
Hamiltonian in Eq.~(\ref{H}), with the hopping matrix elements in 
Eqs.~(\ref{Peierls1}) and (\ref{Peierls2}), has a simple form in the
momentum representation:
\begin{eqnarray}
\mathcal{H}({\bf A})
&=&\sum_{{\bf k}}\left[\epsilon \left({\bf k}-\frac{e{\bf A}(t)}{\hbar c}\right)
-\mu\right] c_{{\bf k}}^{\dagger}c_{{\bf k}}\nonumber\\
&+&\sum_{{\bf k}}
\left[\epsilon^{f}\left({\bf k}-\frac{e{\bf A}(t)}{\hbar c}\right)
-\mu_{f}\right] f_{{\bf k}}^{\dagger}f_{{\bf k}}
\nonumber \\
&+&U\sum_{{\bf p},{\bf k},{\bf q}}
f_{{\bf p}+{\bf q}}^{\dagger}c_{{\bf k}-{\bf q}}^{\dagger}
c_{{\bf k}}f_{{\bf p}},
\label{Hk}
\end{eqnarray}
where the fermionic creation and annihilation operators now create or annihilate
electrons with well-defined momentum.
The free electron energy spectra in Eq.~(\ref{Hk}) satisfy
\begin{eqnarray}
\epsilon\left({\bf k}-\frac{e{\bf A}(t)}{\hbar c}\right)
&=&\epsilon^{f}\left({\bf k}-\frac{e{\bf A}(t)}{\hbar c}\right)
\label{Ek}\\
&=&-2t\lim_{d\rightarrow\infty}
\sum_{j=1}^{d}\cos\left[a\left( {\bf k}_{j}-\frac{e{\bf A}_{j}(t)}{\hbar c}
\right)\right] ,
\nonumber
\end{eqnarray}
for the Hubbard model.  In the case of the Falicov-Kimball model, the
$\epsilon^f$ term vanishes.

We shall consider the spectral moments for the retarded
\begin{equation}
G_{\bf k}^{R}(t_{1},t_{2})
=-i\theta (t_{1}-t_{2})\left\langle
\left\{ c_{{\bf k}}(t_{1}),c_{{\bf k}}^{\dagger}(t_{2})\right\}\right\rangle 
\label{GR}
\end{equation}
and the lesser 
\begin{equation}
G_{\bf k}^{<}(t_{1},t_{2})
=i\left\langle c_{{\bf k}}^{\dagger}(t_{2})c_{{\bf k}}(t_{1})\right\rangle 
\label{Gl}
\end{equation}
Green's functions; the symbol $\{O_1,O_2\}=O_1O_2+O_2O_1$ is the anticommutator 
and the operators $c_{{\bf k}}^{\dagger}(t)$ and $c_{{\bf k}}(t)$
are in the Heisenberg representation, where all time dependence is carried
by the operators and the states are time-independent.
Any Heisenberg representation operator $O_{H}$ is connected
with a corresponding Schr\" odinger representation operator
$O_{S}$ via
\begin{eqnarray}
O_{H}(t)
=\left[ {\mathcal{\bar T}}\exp
\left\{(i/\hbar)\int_{t_{0}}^{t}d{\bar t}\mathcal{H}_{I}({\bar t})\right\}
\right]
e^{(i/\hbar )\mathcal{H}(0)(t-t_{0})}O_{S}
\nonumber \\
\times e^{-(i/\hbar )\mathcal{H}(0)(t-t_{0})}
\left[ \mathcal{T}
\exp\left\{-(i/\hbar)\int_{t_{0}}^{t}d{\bar t}\mathcal{H}_{I}({\bar t})
\right\} \right],
\label{Heisenbergoperator}
\end{eqnarray}
where $\mathcal{H}(0)$ is the time-independent
part of the Hamiltonian [in Eq.~(\ref{H}) with hopping matrix elements given
by their field-free constant values],
and $\mathcal{H}_{I}(t)$ is the time-dependent
part of the Hamiltonian, which includes
the interaction with an external field [but expressed in the interaction
representation as detailed in Eq.~(\ref{interactionoperator}) below]: that is,
we define the time-dependent piece in the Schr\" odinger representation via
$\mathcal{H}_{IS}(t)=\mathcal{H}({\bf A})-
\mathcal{H}(0)$ and then re-express in the interaction 
representation. Note that the interaction
representation operator is defined to be the Schr\" odinger representation
operator evolved under the time-independent Hamiltonian [{\it i.e.}, the
middle three terms in Eq.~(\ref{Heisenbergoperator})].
Hence, the time-dependent piece of the Hamiltonian in the interaction
representation is expressed by
\begin{equation}
\mathcal{H}_{I}(t)
=e^{(i/\hbar )\mathcal{H}(0)(t-t_{0})}\mathcal{H}_{IS}(t)
e^{-(i/\hbar )\mathcal{H}(0)(t-t_{0})},
\label{interactionoperator}
\end{equation}
in terms of the Schr\" odinger operator; in this form, there is the bare time
dependence arising from the time dependence of the fields, plus the time
dependence inherited by the operators, as we go from the Schr\" odinger
representation to the interaction representation.
The symbol $\mathcal{T}$ ($\mathcal{\bar T}$) in Eq.~(\ref{Heisenbergoperator})
is the time-ordering (anti-time-ordering) operator.

We prepare our system to be in equilibrium prior to the field being turned on,
hence the quantum statistical averages in Eqs.~(\ref{GR}) and (\ref{Gl})
are defined with respect to the zero-field (equilibrium) Hamiltonian 
$\mathcal{H}(0)$:
\begin{equation}
\langle (...)\rangle ={\rm Tr}\left[ e^{-\beta \mathcal{H}(0)}(...)\right] /
\mathcal{Z},
\label{average}
\end{equation} 
where the partition function satisfies
\begin{equation}
\mathcal{Z}={\rm Tr}\left[ e^{-\beta \mathcal{H}(0)}\right] ,
\label{Z}
\end{equation}
and $\beta$ is the inverse temperature of the original equilibrium distribution.
As was already mentioned in Section I, the retarded
and the lesser Green's functions form an independent Green's
function basis. Any other Green's function can be expressed
in terms of these two functions. This is in contrast to the equilibrium
case, where only one Green's function is independent (because the Fermi-Dirac
distribution function is determined by the equilibrium condition).

Calculations in the nonequilibrium case are complicated
by the fact that the Green's functions in Eqs.~(\ref{GR}) and (\ref{Gl})
are functions of two time variables, contrary to the equilibrium
case, where they only depend on the relative time difference (because the
equilibrium system is time-translation invariant).
It is convenient to transform the two-time dependence of the Green's functions
from $t_1$ and $t_2$ to Wigner coordinates, which use the average time
$T=(t_{1}+t_{2})/2$ and the relative time $t=t_{1}-t_{2}$ (do not confuse
the average time $T$ with the temperature $1/\beta$). 
Next, the relative time dependence is Fourier transformed to a frequency,
and the additional (average) time evolution
of different quantities is described by the average time
coordinate $T$; in equilibrium, there is no $T$ dependence. For example, 
the spectral function for the retarded and the lesser Green's functions
can be defined as
\begin{eqnarray}
A_{{\bf k}}^{R,<}(T,\omega)
=\int_{-\infty}^{\infty}dt e^{i\omega t}
\left( \eta\frac{1}{\pi}\right)
{\rm Im}G_{{\bf k}}^{R,<}(T,t),
\label{AR<}
\end{eqnarray}
where $G_{{\bf k}}^{R,<}(T,t)$ is the respective Green's function from 
Eq.~(\ref{GR}) or (\ref{Gl}) with $t_1=T+t/2$ and $t_2=T-t/2$; $\eta$ is equal to
$-1$ for the retarded Green's function and $+1$ for the lesser Green's
function so that the spectral functions are nonnegative in equilibrium.
In general, the spectral function depends on the average time, because the
system no longer has time-translation invariance when a field is turned on
at a specific time.

We define the $n$th moment of the retarded and lesser spectral function 
[in Eq.~(\ref{AR<})] $\mu_{n}^{R,<}$ to be
\begin{eqnarray}
\mu_{n}^{R,<}({\bf k},T)&=&\int_{-\infty}^{\infty}d\omega \omega^{n}
A_{{\bf k}}^{R,<}(T,\omega)\nonumber\\
&=&\int_{-\infty}^{\infty}d\omega \omega^{n}
\left( \eta\frac{1}{\pi}\right)
{\rm Im}G_{{\bf k}}^{R,<}(T,\omega)
\nonumber \\
&=&\int_{-\infty}^{\infty}d\omega \omega^{n}
\left( \eta\frac{1}{\pi}\right)\nonumber\\
&~&\times
{\rm Im}\int_{-\infty}^{\infty}dt e^{i\omega t}
G_{{\bf k}}^{R,<}(T,t).
\label{munR<}
\end{eqnarray}
It is easy to show that this expression is equivalent to
\begin{eqnarray}
\mu_{n}^{R,<}({\bf k},T)&=&
\eta\frac{1}{\pi}
\int_{-\infty}^{\infty}d\omega\nonumber\\
&~&\times
{\rm Im}\int_{-\infty}^{\infty}dt e^{i\omega t}
i^{n}\frac{\partial^{n}}{\partial t^{n}}
G_{{\bf k}}^{R,<}(T,t).
\label{munR<2}
\end{eqnarray}
Begin by noting that
\begin{equation}
\frac{\partial^{n}}{(-i)^{n}\partial t^{n}}G_{{\bf k}}^{R,<}(T,t)
=\int_{-\infty}^{\infty}\frac{d\omega}{2\pi} e^{-i\omega t}
\omega^{n}G_{{\bf k}}^{R,<}(T,\omega ),
\label{munR<3}
\end{equation}
so that
\begin{equation}
\omega^{n}G_{{\bf k}}^{R,<}(T,\omega )=
\int_{-\infty}^{\infty}dt e^{i\omega t}
\frac{\partial^{n}}{(-i)^{n}\partial t^{n}}G_{{\bf k}}^{R,<}(T,t).
\label{munR<4}
\end{equation}
Substituting Eq.~(\ref{munR<4}) into the first line of 
Eq.~(\ref{munR<}) then yields Eq.~(\ref{munR<2}). 
Before proceeding with the evaluation of  
analytical expressions for the spectral
moments from Eq.~(\ref{munR<2}), we note that the integration
over frequency in Eq.~(\ref{munR<2}) can be evaluated,
yielding the following expression, which connects the spectral
moments to the derivative 
of the Green's function with respect to relative time $t$ at zero relative time:
\begin{equation}
\mu_{n}^{R,<}({\bf k},T)=
\eta2{\rm Im}
\left[
i^{n}\frac{\partial^{n}}{\partial t^{n}}
G_{{\bf k}}^{R,<}(T,t)
\right]_{t=0^{+}} .
\label{munR<5}
\end{equation}
This formula assumes that the Green's function is a differentiable
function, which is true in most cases of interest.
Despite the fact that the expressions in Eqs.~(\ref{munR<2}) and (\ref{munR<5})
are formally equivalent, it is preferable to use one or the other
in specific cases. 

In the case of the retarded Green's function, the well-known expression
for the zeroth spectral moment can be found from Eqs.~(\ref{munR<5}) and 
(\ref{GR}):
\begin{eqnarray}
&~&\mu_{0}^{R}({\bf k},T)=-2{\rm Im}
\left[G_{{\bf k}}^{R}(T,t)\right]_{t=0^{+}}\nonumber\\
&=&-2{\rm Im}\left[ 
-i\theta (t)
\left\langle\left\{ c_{{\bf k}}\left( T+\frac{t}{2}\right) ,
c_{{\bf k}}^{\dagger}\left( T-\frac{t}{2}\right)\right\}
\right\rangle 
\right]_{t=0^{+}}
\nonumber \\
&=&\left\langle\left\{ c_{{\bf k}}(T),
c_{{\bf k}}^{\dagger}(T)\right\}\right\rangle =1.
\label{mu0R}
\end{eqnarray}
In the derivation of Eq.~(\ref{mu0R}), we used the
anticommutation relation for Heisenberg operators
and the fact that the theta function is equal to $1/2$
when its argument is equal to zero.

It is more convenient to use Eqs.~(\ref{munR<2}) and (\ref{GR}) to evaluate
the expression for the first moment of $G^{R}$:
\begin{eqnarray}
&~&\mu_{1}^{R}({\bf k},T)=
\left( -\frac{1}{\pi}\right)
\int_{-\infty}^{\infty}d\omega
{\rm Im}\int_{-\infty}^{\infty} dt e^{i\omega t}\nonumber\\
&\times&\frac{\partial}{\partial t}
\left[
\theta (t)
\left\langle\left\{ c_{{\bf k}}\left(T+\frac{t}{2}\right),
c_{{\bf k}}^{\dagger}\left( T-\frac{t}{2}\right) \right\}\right\rangle 
\right] .
\label{mu1Rderiv}
\end{eqnarray}
Taking the time derivative in Eq.~(\ref{mu1Rderiv}) 
gives
\begin{eqnarray}
\mu_{1}^{R}({\bf k},T)&=&
-\frac{1}{\pi}\int_{-\infty}^{\infty}d\omega
{\rm Im} \int_{-\infty}^{\infty} dt e^{i\omega t}
\delta (t)\nonumber\\
&\times& \left\langle\left\{ c_{{\bf k}}\left( T+\frac{t}{2}\right) ,
c_{{\bf k}}^{\dagger}\left( T-\frac{t}{2}\right)\right\}
\right\rangle 
\nonumber \\
&+&\frac{1}{\pi}
\int_{-\infty}^{\infty}d\omega
{\rm Im} \int_{-\infty}^{\infty} dt e^{i\omega t}
i\theta (t)\nonumber\\
&\times&\left[
\left\langle\left\{ i\frac{\partial}{\partial t}
c_{{\bf k}}\left( T+\frac{t}{2}\right)
                ,c_{{\bf k}}^{\dagger}\left( T-\frac{t}{2}\right) \right\} 
\right\rangle
\right.
\nonumber \\
&+&\left.\left\langle\left\{ c_{{\bf k}}\left( T+\frac{t}{2}\right)
    ,i\frac{\partial}{\partial t}c_{{\bf k}}^{\dagger}\left( T-\frac{t}{2}\right) \right\}
\right\rangle
\right] .
\label{mu1Rderiv2}
\end{eqnarray}
The first term in Eq.~(\ref{mu1Rderiv2}) is equal to zero,
because the integral over time is equal to $1$, therefore
its imaginary part vanishes.
The second term in Eq.~(\ref{mu1Rderiv2}) can be simplified by performing
an integration over $\omega$ and replacing the time derivatives
of the operators by their commutators with the Hamiltonian,
according to the Heisenberg equation of motion
$i\partial O(t)/\partial t \
=[O(t),\mathcal{H}(t)]$, where $\mathcal{H}(t)$ is the total
Hamiltonian including the effects of the time-dependent field. This yields
\begin{eqnarray}
\mu_{1}^{R}({\bf k},T)&=&
\frac{1}{2}{\rm Re}
\left(
\left\langle\left\{ [c_{{\bf k}}(T),\mathcal{H}(T)],c_{{\bf k}}^{\dagger}(T)
\right\} \right\rangle\right .\nonumber\\
&-&\left.
\left\langle\left\{ c_{{\bf k}}(T),[c_{{\bf k}}^{\dagger}(T),\mathcal{H}(T)]
\right\}\right\rangle\right) .
\label{mu1Rder3}
\end{eqnarray}
Evaluation of the commutators of the Fermi-operators
with the Hamiltonian and the subsequent anticommutators in Eq.~(\ref{Hk}) 
gives the following expression
for the first spectral moment of the retarded Green's function
\begin{eqnarray}
\mu_{1}^{R}({\bf k},T)=
\epsilon \left({\bf k}-\frac{e{\bf A}(T)}{\hbar c}\right)
-\mu+Un_{f},
\label{mu1R1}
\end{eqnarray}
where
\begin{equation}
n_{f}=\sum_{{\bf k}}\langle f_{{\bf k}}^{\dagger}(T)f_{{\bf k}}(T)\rangle
\label{nf}
\end{equation}
is the average  
number of $f$ ($c_{\downarrow}$)-electrons
in the system; this number of electrons does not depend on the average or
the relative time, because the total electron number for each species of electron
is conserved.

Similarly, the expression for the
second moment of the retarded Green's function can be found from
Eqs.~(\ref{munR<2}) and (\ref{GR}):
\begin{eqnarray}
\mu_{2}^{R}({\bf k},T)&=&
\frac{1}{4}{\rm Re}
\left(
\left\langle\left\{ [[c_{{\bf k}}(T),\mathcal{H}(T)],\mathcal{H}(T)],
c_{{\bf k}}^{\dagger}(T)
\right\}\right\rangle
\right.
\nonumber \\
&-&2\left\langle\left\{ [c_{{\bf k}}(T),\mathcal{H}(T)],[c_{{\bf k}}^{\dagger}(T),
\mathcal{H}(T)]
\right\}\right\rangle\nonumber\\
&+&\left. \left\langle\left\{ c_{{\bf k}}(T),[[c_{{\bf k}}^{\dagger}(T),
\mathcal{H}(T)],\mathcal{H}(T)]
\right\}\right\rangle
\right) .
\label{mu2Rderiv2}
\end{eqnarray}
Details of the derivation are presented in the Appendix.

Evaluating the commutators and anticommutators in Eq.~(\ref{mu2Rderiv2}) gives
\begin{eqnarray}
\mu_{2}^{R}({\bf k},T)
&=&\left[\epsilon \left({\bf k}-\frac{e{\bf A}(T)}{\hbar c}\right)
-\mu\right]^{2}\label{mu2R}
\\
&+&2U\left[\epsilon \left({\bf k}-\frac{e{\bf A}(T)}{\hbar c}\right)
-\mu\right]n_{f}+U^{2}n_{f}.
\nonumber   
\end{eqnarray}

The moments of the local retarded Green's function 
${\tilde \mu}_{n}^{R}(T)$ are obtained
by summing the corresponding 
spectral moment functions $\mu_{n}^{R}({\bf k},T)$ over ${\bf k}$
\begin{equation}
{\tilde \mu}_{n}^{R}(T)=\sum_{{\bf k}}\mu_{n}^{R}({\bf k},T).
\label{mulocal}
\end{equation}
Performing the summations for Eqs.~(\ref{mu1R1}) and (\ref{mu2R}) yields
the following local moments:
\begin{eqnarray}
{\tilde \mu}_{1}^{R}(T)&=&-\mu+Un_{f} ;
\label{mu1Rdensity}\\
{\tilde \mu}_{2}^{R}(T)
&=&\frac{1}{2}+\mu^{2}-2U\mu n_{f}+U^{2}n_{f} .
\label{mu2Rdensity}
\end{eqnarray}
These results coincide with those derived previously for the Hubbard model
in the equilibrium case\cite{white}.
Since the hopping matrix is always chosen to be traceless in our models,
the sum of the energy $\sum_{{\bf k}}\epsilon ({\bf k})$ 
in Eqs.~(\ref{mu1R1}) and (\ref{mu2R}) is equal to zero.
The expression for the zeroth local moment ${\tilde \mu}_{0}^{R}$
has the same form as the expression for the zeroth 
spectral moment in Eq.~(\ref{mu0R}), since the zeroth spectral moment
is momentum-independent.  Hence, the zeroth and the first two local
moments of the retarded Green's function in an
arbitrary external time-dependent homogeneous electric
field are all time-independent!  This is a nontrivial result because
the retarded Green's function strongly depends on the average time.
In particular, the retarded Green's function is an oscillating function
of time\cite{Turkowski} when $U=0$. Furthermore, the moments do not depend on the
electric field at half-filling because the chemical potential is not changed by
the field! It isn't obvious whether the chemical potential would be changed
by the field off of half-filling.

In the case of half-filling, where $n_c=n_{f}=1/2$ and $\mu =U/2$,
the expressions in Eqs.~(\ref{mu0R}), (\ref{mu1Rdensity}) and (\ref{mu2Rdensity}) 
acquire an even simpler form:
\begin{eqnarray}
{\tilde \mu}_{0}^{R}(T)&=&1,
\label{mu0Rdensityhalffilling}\\
{\tilde \mu}_{1}^{R}(T)&=&0,
\label{mu1Rdensityhalffilling}\\
{\tilde \mu}_{2}^{R}(T)&=&\frac{1}{2}+\frac{U^{2}}{4}.
\label{mu2Rdensityhalffilling}
\end{eqnarray}

If one examines the moments for gauge-invariant Green's
functions,\cite{gauge_inv}~then the local moments are unchanged, and the
spectral function moments are modified by a time-dependent shift of
the momentum wavevector.  We don't include those formulas here, because they
just involve such a simple shift.

The corresponding moments of the lesser Green's functions are found by a
similar analysis.  Using Eqs.~(\ref{munR<5}) and (\ref{Gl}) we find
\begin{eqnarray}
\mu_{0}^{<}({\bf k},T)&=&2n_{c}({\bf k},T) ,
\label{mu0<}\\
\mu_{1}^{<}({\bf k},T)&=&-
{\rm Re}
\left(
\left\langle[c_{{\bf k}}^{\dagger}(T),\mathcal{H}(T)]
c_{{\bf k}}(T)\right\rangle\right.
\nonumber\\
&-&\left. \left\langle c_{{\bf k}}^{\dagger}(T)[c_{{\bf k}}(T),\mathcal{H}(T)]
\right\rangle \right) ,
\label{mu1<deriv}\\
\mu_{2}^{<}({\bf k},T)&=&\frac{1}{2}{\rm Re}
\left(
\left\langle [[c_{{\bf k}}^{\dagger}(T),\mathcal{H}(T)],\mathcal{H}(T)]
c_{{\bf k}}(T)\right\rangle\right.
\nonumber\\
&-&\left.2\left\langle [c_{{\bf k}}^{\dagger}(T),\mathcal{H}(T)]
[c_{{\bf k}}(T),\mathcal{H}(T)]\right\rangle
\right.
\nonumber \\
&+&\left. \left\langle c_{{\bf k}}^{\dagger}(T)[[c_{{\bf k}}^{\dagger}(T),
\mathcal{H}(T)],\mathcal{H}(T)]
\right\rangle
\right) ,
\label{mu2<deriv}
\end{eqnarray}
where 
\begin{equation}
n_{c}({\bf k},T)=\left\langle c_{{\bf k}}^{\dagger}(T)c_{{\bf k}}(T)
\right\rangle
\label{MDF}
\end{equation}
is the momentum distribution function for the $c$~$(c_{\uparrow})$-electrons.
Note that a commutator term depending on the derivative of the Hamiltonian
with respect to time can be shown to cancel, so it is not included in the
second moment expression above.
Evaluation of the commutators of the operators with the Hamiltonian 
[in Eq.~(\ref{Hk})] in Eqs.~(\ref{mu1<deriv}) and (\ref{mu2<deriv})
gives
\begin{eqnarray}
\mu_{1}^{<}({\bf k},T)
=2 \left[\epsilon \left({\bf k}-\frac{e{\bf A}(T)}{\hbar c}\right)
-\mu\right]n_{c}({\bf k},T)
\nonumber \\
+U\sum_{{\bf p},{\bf q}}\left(\left\langle f_{{\bf p}+{\bf q}}^{\dagger}
                         c_{{\bf k}-{\bf q}}^{\dagger}
                         c_{{\bf k}}f_{{\bf p}}\right\rangle
                        +\left\langle f_{{\bf p}+{\bf q}}^{\dagger}
                         c_{{\bf k}}^{\dagger}
                         c_{{\bf k}+{\bf q}}f_{{\bf p}}\right\rangle\right) .
\label{mu1<}
\end{eqnarray}
and
\begin{widetext}
\begin{eqnarray*}
\mu_{2}^{<}({\bf k},T)
&=&2\left[\epsilon \left({\bf k}-\frac{e{\bf A}(T)}{\hbar c}\right)
-\mu\right]^{2}n_{c}({\bf k},T)
+\frac{3}{2}U\left[\epsilon \left({\bf k}-\frac{e{\bf A}(T)}{\hbar c}\right)
-\mu\right]
\sum_{{\bf p},{\bf q}}\left(\left\langle f_{{\bf p}+{\bf q}}^{\dagger}
                        c_{{\bf k}-{\bf q}}^{\dagger}
                        c_{{\bf k}}f_{{\bf p}}\right\rangle
                      +\left\langle f_{{\bf p}+{\bf q}}^{\dagger}
                        c_{{\bf k}}^{\dagger}
                        c_{{\bf k}+{\bf q}}f_{{\bf p}}\right\rangle\right) 
\nonumber \\
&+&\frac{1}{2}U\sum_{{\bf p},{\bf q}}
\left[\epsilon \left({\bf k}-{\bf q}-\frac{e{\bf A}(T)}{\hbar c}\right)
-\mu\right] \left\langle 
f_{{\bf p}+{\bf q}}^{\dagger}c_{{\bf k}-{\bf q}}^{\dagger}
             c_{{\bf k}}f_{{\bf p}}\right\rangle
+\frac{1}{2}U\sum_{{\bf p},{\bf q}}
\left[\epsilon \left({\bf k}+{\bf q}-\frac{e{\bf A}(T)}{\hbar c}\right)
-\mu\right] \left\langle f_{{\bf p}+{\bf q}}^{\dagger}c_{{\bf k}}^{\dagger}
             c_{{\bf k}+{\bf q}}f_{{\bf p}}\right\rangle
\nonumber \\
&+&\frac{1}{2}U\sum_{{\bf p},{\bf q}}
\epsilon^{f}\left({\bf p}+{\bf q}-\frac{e{\bf A}(T)}{\hbar c}\right)
\left[ \left\langle f_{{\bf p}+{\bf q}}^{\dagger}c_{{\bf k}}^{\dagger}
        c_{{\bf k}+{\bf q}}f_{{\bf p}}\right\rangle
- \left\langle f_{{\bf p}+{\bf q}}^{\dagger}c_{{\bf k}-{\bf q}}^{\dagger}
   c_{{\bf k}}f_{{\bf p}}\right\rangle
\right]
\nonumber \\
&-&\frac{1}{2}U\sum_{{\bf p},{\bf q}}
\epsilon^{f}\left({\bf p}-\frac{e{\bf A}(T)}{\hbar c}\right)
\left[ \left\langle f_{{\bf p}+{\bf q}}^{\dagger}c_{{\bf k}}^{\dagger}
        c_{{\bf k}+{\bf q}}f_{{\bf p}}\right\rangle
- \left\langle f_{{\bf p}+{\bf q}}^{\dagger}c_{{\bf k}-{\bf q}}^{\dagger}
   c_{{\bf k}}f_{{\bf p}}\right\rangle
\right]
\nonumber \\
&+&\frac{1}{2}U^{2}\sum_{{\bf p},{\bf q},{\bf P},{\bf Q}}
\left[ \left\langle f_{{\bf p}+{\bf q}}^{\dagger}f_{{\bf p}}
        f_{{\bf P}+{\bf Q}}^{\dagger}f_{{\bf P}}
        c_{{\bf k}-{\bf q}-{\bf Q}}^{\dagger}c_{{\bf k}}\right\rangle
+2\left\langle f_{{\bf p}+{\bf q}}^{\dagger}f_{{\bf p}}
        f_{{\bf P}+{\bf Q}}^{\dagger}f_{{\bf P}}
        c_{{\bf k}-{\bf q}}^{\dagger}c_{{\bf k}+{\bf Q}}\right\rangle
\right.
\nonumber \\
&+&\left. \left\langle f_{{\bf p}+{\bf q}}^{\dagger}f_{{\bf p}}
        f_{{\bf P}+{\bf Q}}^{\dagger}f_{{\bf P}}
        c_{{\bf k}}^{\dagger}c_{{\bf k}+{\bf q}+{\bf Q}}\right\rangle 
\right ]. \label{mu2<}
\end{eqnarray*}
\end{widetext}
\addtocounter{equation}{1}
In Eq.~(\ref{mu2<}), we have suppressed the time label $T$ corresponding
to the time at which all operators are evaluated. We continue to suppress
this time label in some equations below; this should not cause any confusion.

The expressions in Eqs.~(\ref{mu0<})--(\ref{mu2<}) for the lesser spectral 
moments are more complicated than the corresponding retarded moments.
However, they simplify in the case of the local Green's function, where we find
\begin{eqnarray}
{\tilde \mu}_{0}^{<}(T)&=&2n_{c},
\label{mu0<density}\\
{\tilde \mu}_{1}^{<}(T)
&=&2\sum_{{\bf k}}
\left[
\epsilon \left({\bf k}-\frac{e{\bf A}(T)}{\hbar c}\right)-\mu
\right]n_{c}({\bf k},T)\nonumber\\
&+&2U\sum_{i}\left\langle f_{i}^{\dagger}f_{i}c_{i}^{\dagger}c_{i}\right\rangle ,
\label{mu1<density}\\
{\tilde \mu}_{2}^{<}(T)
&=&2\sum_{{\bf k}}
\left[
\epsilon \left({\bf k}-\frac{e{\bf A}(T)}{\hbar c}\right)-\mu
\right]^{2}n_{c}({\bf k},T)
\nonumber \\
&+&2U\sum_{{\bf k},{\bf p},{\bf q}}
\left[ 
\epsilon\left({\bf k}-\frac{e{\bf A}(T)}{\hbar c}\right)\right.\nonumber\\
&~&\quad +\left. \epsilon\left({\bf k}-{\bf q}-\frac{e{\bf A}(T)}{\hbar c}\right)
\right]
\left\langle f_{{\bf p}+{\bf q}}^{\dagger}f_{{\bf p}}
        c_{{\bf k}-{\bf q}}^{\dagger}c_{{\bf k}}\right\rangle
\nonumber \\
&-&2U(2\mu-U)\sum_{i}\left\langle f_{i}^{\dagger}f_{i}c_{i}^{\dagger}c_{i}
\right\rangle ,
\label{mu2<density}
\end{eqnarray}
with
\begin{equation}
n_{c}=\sum_{{\bf k}}n_{c}({\bf k},T)
\label{nc}
\end{equation}
being the time-independent particle density of the
$c$~($c_{\uparrow}$)-electrons.
In order to save space, and make the equations more transparent,
we use a mixed real-space/momentum-space representation
for the operators in Eqs.~(\ref{mu1<density}) and (\ref{mu2<density}).
Note that the first moment involves one correlation function and the
second moment involves two correlation functions. Note further that
the value of the first moment in Eq.~(\ref{mu1<density}) is equal
to twice the average value of the Hamiltonian.
The last term in Eq.~(\ref{mu2<density}) is equal to zero in the case
of half-filling ($\mu=U/2$). 
The second two moments of the lesser Green's functions
do appear to depend both on the average time and the electric field
(although our empirical evidence at half filling suggests the second moment
may be independent of average time---see the numerical results below).

There is an interesting observation that can be made about the first moment,
and its relation to the current driven by the electric field
and the phenomenon of Bloch oscillations.
In the limit where $U$ is small, one can evaluate the correlation function
in Eq.~(\ref{mu1<density})
via a mean-field theory decoupling ($\langle f_i^\dagger f_ic_i^\dagger c_i\rangle
\approx \langle f_i^\dagger f_i\rangle \langle c_i^\dagger c_i\rangle$).  
For example, at half filling, the first moment will be equal to twice the average
kinetic energy (including the shift by the vector potential needed to
construct the actual kinetic energy from the bandstructure) plus a correction 
of order $U^2$.
If the current oscillates, we expect the average kinetic energy to oscillate
as well.  Hence, for small $U$ there is a correlation between oscillations
in the first moment of the local Green's function and oscillations of the current.

The correlation function $\left\langle f^\dagger_if_ic^\dagger_ic_i\right\rangle$
that appears in Eqs.~(\ref{mu1<density}) and (\ref{mu2<density}) can be
determined for the Falicov-Kimball or Hubbard model via the equation of motion,
because it is related to the total energy of the Hamiltonian.  To show how
this works, we first provide the derivation for the equilibrium case using an
imaginary-time formalism. Begin with the definition of the Green's function in 
real space
\begin{equation}
G_{ij}(\tau)=-\left \langle \mathcal{T}c_i(\tau)c_j^\dagger(0)\right\rangle,
\label{eq: g_tau}
\end{equation}
with a similar result for the spin-down electrons in the Hubbard model. Here we
have $c_i(\tau)=\exp(\mathcal{H}\tau)c_i(0)\exp(-\mathcal{H}\tau)$.
Taking the imaginary-time derivative of the local Green's function gives
\begin{eqnarray}
\partial_\tau G_{ii}(\tau)&=&-\delta(\tau)-\left\langle\mathcal{T}[\mathcal{H},
c_i(\tau)]c^\dagger_i(0)\right\rangle\nonumber\\
&=&-\delta(\tau)+t\sum_\delta G_{i+\delta i}(\tau)+\mu G_{ii}(\tau)\nonumber\\
&+&
U\left\langle\mathcal{T}f_i^\dagger f_ic_i(\tau)c_i^\dagger(0)\right\rangle,
\label{eq: eom}
\end{eqnarray}
where the symbol $\delta$ denotes the translation vector to a nearest-neighbor
site and $i+\delta$ as a subscript refers to the lattice site that is the
nearest neighbor of site $i$ translated by the nearest-neighbor translation
vector $\delta$.  Hence we determine the correlation function via
\begin{eqnarray}
U\left\langle\mathcal{T}f_i^\dagger f_ic_i^\dagger c_i\right\rangle
&=&\lim_{\tau\rightarrow 0^-}\left [
-\partial_\tau G_{ii}(\tau)\right. \label{eq: corr_fun}\\
&+&\left. t\sum_\delta G_{i+\delta i}(\tau)+\mu G_{ii}(\tau)
-\delta(\tau)\right ].
\nonumber             
\end{eqnarray}
Using the Matsubara frequency representation 
\begin{equation}
G(i\omega_n)=G_n=\int_0^\beta
d\tau e^{i\omega_n\tau}G(\tau),
\label{eq: mats}
\end{equation}
with $i\omega_n=i\pi(2n+1)/\beta$ the fermionic Matsubara frequency, allows us
to determine a simple expression for the correlation function.  Note that
$G(\tau)=(1/\beta)\sum_n\exp(-i\omega_n\tau)G_n$ and that the Green's
function in momentum space satisfies 
$G_n({\bf k})=1/[i\omega_n+\mu-\Sigma_n({\bf k})-\epsilon({\bf k})]$, to find 
that the correlation function simplifies to
\begin{equation}
\left\langle\mathcal{T}f_i^\dagger f_ic_ic_i^\dagger\right\rangle
=\frac{1}{\beta U}\sum_n\sum_{\bf k}\frac{\Sigma_n({\bf k})}
{i\omega_n+\mu-\Sigma_n({\bf k})-\epsilon({\bf k})}.
\label{eq: corr_fun2}
\end{equation}
In the limit of infinite dimensions, the self-energy is a local function,
and hence has no momentum dependence. Then the sum over momentum can be
performed by changing from a sum over momentum to an integral over the
noninteracting density of states. This produces the local Green's function, and
we are left with the final form for DMFT:
\begin{equation}
\left\langle\mathcal{T}f_i^\dagger f_ic_i^\dagger c_i\right\rangle
=\frac{1}{\beta U}\sum_n\Sigma_nG_n.
\label{eq: corr_fun3}
\end{equation}
In numerical calculations, it is more convenient to evaluate the summation
in Eq.~(\ref{eq: corr_fun3}) via the formally equivalent expression
with the Hartree-Fock contribution to the self-energy removed
\begin{equation}
\left\langle\mathcal{T}f_i^\dagger f_ic_i^\dagger c_i\right\rangle
=\langle f^\dagger f\rangle \langle c^\dagger c\rangle+
\frac{1}{\beta U}\sum_n[\Sigma_n-U\langle f^\dagger f\rangle]G_n,
\label{eq: corr_fun4}
\end{equation}
because the Matsubara summation converges faster.

In the nonequilibrium case, one can perform a similar analysis, but now one
has to work with the nonequilibrium Green's functions as functions of real
time variables.  Using the standard equations of motion, and definitions for
nonequilibrium Green's functions, one can show, after some significant
algebra, that the correlation function can be expressed as
\begin{eqnarray}
U\left\langle f_i^\dagger f_ic_i^\dagger c_i\right\rangle&=&-i
\sum_{\bf k}\left [ i\frac{\partial}{\partial t_1}+\mu-\epsilon\left({\bf k}-
\frac{e{\bf A}(t_1)}{\hbar c}\right)\right ] \nonumber\\
&\times&G^<_{\bf k}(t_1,t_2)\Biggr |_{t_2=t_1}
\nonumber\\
&=&-i\sum_{\bf k}\int dt\left [ \Sigma^R_{\bf k}(t_1,t)G^<_{\bf k}(t,t_1)\right.
\nonumber\\
&+&\left. \Sigma^<_{\bf k}(t_1,t)G^A_{\bf k}(t,t_1)\right ],
\label{eq: corr_noneq}
\end{eqnarray}
where the final result is written in terms of retarded, lesser and advanced
Green's functions and self-energies in the presence of the field. In the DMFT
limit, the self-energies have no momentum dependence. This expression
appears like it can have average time dependence, but we cannot say that
it definitely does, because there could be a cancellation of the time
dependence.

As a check of Eq.~(\ref{eq: corr_noneq}), we evaluate it in equilibrium, to show
it yields the same result as Eq.~(\ref{eq: corr_fun3}). When we are in 
equilibrium, the correlation function is independent of time and the
Green's functions and self-energies depend only on the time difference
of their two time variables.  Hence, we can perform a Fourier transform
by using the convolution theorem to transform Eq.~(\ref{eq: corr_noneq}) into
\begin{eqnarray}
U\left\langle f_i^\dagger f_ic_i^\dagger c_i\right\rangle&=&
-\frac{i}{2\pi}\sum_{\bf k}\int d\omega\left [
\Sigma^R_{\bf k}(\omega)G^<_{\bf k}(\omega)\right. \nonumber\\
&+&\left. \Sigma^<_{\bf k}(\omega)
G^A_{\bf k}(\omega)\right ].
\label{eq: corr_fun_eq}
\end{eqnarray}
Using the fact that the lesser functions satisfy
$G^<_{\bf k}=-2if(\omega){\rm Im}G^R_{\bf k}(\omega)$ and
$\Sigma^<_{\bf k}=-2if(\omega){\rm Im}\Sigma^R_{\bf k}(\omega)$ in
equilibrium, allows us to transform Eq.~(\ref{eq: corr_fun_eq}) into
\begin{equation}
U\left\langle f_i^\dagger f_ic_i^\dagger c_i\right\rangle=
-\frac{1}{\pi}\sum_{\bf k}\int d\omega f(\omega){\rm Im}\left [
\Sigma_{\bf k}^R(\omega)G^R_{\bf k}(\omega)\right ],
\label{eq: corr_fun_eq2}
\end{equation}
which is equal to the analytic continuation of Eq.~(\ref{eq: corr_fun3})
from the imaginary axis to the real axis.

The correlation function $\left\langle f_{{\bf p}+{\bf q}}^{\dagger}f_{{\bf p}}
c_{{\bf k}-{\bf q}}^{\dagger}c_{{\bf k}}\right\rangle$ which enters
Eq.~(\ref{mu2<density}) is more complicated to evaluate, because it cannot
be expressed in terms of a simple equation of motion for the single-particle
Green's functions.  Because of it's complex nature, we will evaluate it
only for the Falicov-Kimball model in equilibrium.  While it is certainly true
that it can be evaluated for the Falicov-Kimball model in the presence of 
a field, some of the formal details become quite complicated, and take us
away from the main theme of this work, so we do not perform such an analysis
here.

Our starting point, then, is the operator average
\begin{equation}
\sum_{\bf k,p,q}\left [ \epsilon({\bf k})+\epsilon({\bf k}-{\bf q})\right ]
\left\langle f_{{\bf p}+{\bf q}}^{\dagger}f_{{\bf p}}
c_{{\bf k}-{\bf q}}^{\dagger}c_{{\bf k}}\right\rangle,
\label{eq: 2corr_fun}
\end{equation}
where we have set the vector potential ${\bf A}$ equal to zero. We use a Fourier
transform to express the localized electrons in terms of their real-space
operators.  Then Elitzur's theorem\cite{elitzur} ensures that the
operator expectation value vanishes if the two localized
electrons are not at the same
lattice site ({\it i.e.}, there is no spontaneous hybridization in the
Falicov-Kimball model for nonzero temperature). This allows us to perform the
summation over the momentum variable ${\bf p}$ and gives us
\begin{equation}
\sum_{\bf k,q}\sum_i\left [ \epsilon({\bf k})+\epsilon({\bf k}-{\bf q})\right ]
e^{i{\bf q}\cdot{\bf R_i}}
\left\langle f_i^{\dagger}f_i
c_{{\bf k}-{\bf q}}^{\dagger}c_{{\bf k}}\right\rangle.
\label{eq: 2corr_fun2}
\end{equation}
Next, we express the bandstructure in terms of the summation over nearest
neighbor translation vectors $\delta$: $\epsilon({\bf k})=-t^*\sum_\delta
\exp[i{\bf k}\cdot\delta]/\sqrt{d}$, and introduce Fourier transforms for
the itinerant electrons to real space. This allows us to sum over the remaining
momenta, yielding
\begin{equation}
-\frac{t^*}{\sqrt{d}}\sum_{i\delta}\left [ 
\left\langle f_i^{\dagger}f_i c_i^{\dagger}c_{i+\delta}\right\rangle+
\left\langle f_i^{\dagger}f_i c_{i+\delta}^{\dagger}c_i\right\rangle\right ].
\label{eq: 2corr_fun3}
\end{equation}
The statistical averages in Eq.~(\ref{eq: 2corr_fun3}) have already been
evaluated.\cite{thermal}~The procedure is to imagine adding a small 
field $-\sum_i h_i f^\dagger_if_i$ to the Hamiltonian, and evaluate
the expectation value with the localized particle number via a derivative
with respect to the field strength $h_i$ (then set the field to zero to
evaluate the average). This gives
\begin{eqnarray}
&~&-\frac{t^*}{\sqrt{d}}\sum_{i\delta}\left [ \frac{1}{\beta}
\frac{\partial}{\partial h_i}+\langle w_i\rangle\right ]
\left [ \left\langle c_i^{\dagger}c_{i+\delta}\right\rangle+
\left\langle c_{i+\delta}^{\dagger}c_i\right\rangle\right ]\nonumber\\
&=&-\frac{t^*}{\sqrt{d}}\sum_{i\delta}\left [ \frac{1}{\beta}
\frac{\partial}{\partial h_i}+\langle w_i\rangle\right ]\nonumber\\
&~&\times
\left [ G_{i+\delta i}(\tau\rightarrow 0^-)+G_{ii+\delta}(\tau\rightarrow 0^-)
\right ].
\label{eq: 2corr_fun4}
\end{eqnarray}
Now we follow the derivation in Ref.~\onlinecite{thermal}, which evaluates the
derivatives from the following:
\begin{eqnarray}
&~&\left [ \frac{1}{\beta}
\frac{\partial}{\partial h_i}+\langle w_i\rangle\right ]
G_{i+\delta i}(\tau\rightarrow 0^-)\label{eq: 2corr_fun5}\\
&~&=\frac{1}{\beta}\sum_n\sum_{jk}
G_{i+\delta j}\left [ -\frac{1}{\beta}
\frac{\partial}{\partial h_i}+\langle w_i\rangle\right ]
G^{-1}_{jk}(i\omega_n)G_{ki}(i\omega_n)\nonumber\\
&~&=\frac{1}{\beta}\sum_n\left [
G_{i+\delta i} \frac{1}{\beta}
\frac{\partial}{\partial h}\Sigma(i\omega_n)
G_{ii}(i\omega_n)+\langle f^\dagger f\rangle G_{ii}(i\omega_n)\right ]
\nonumber
\end{eqnarray}
where the derivative with respect to the field $h$ acts on the local
self-energy.  After some long and complicated algebra, that derivative
can be determined, which yields our final result for the correlation function
\begin{eqnarray}
&~&
\frac{2}{\beta}\sum_n\sum_{\bf k}\frac{\epsilon({\bf k})}{U}G_{\bf k}(i\omega_n)
\Sigma(i\omega_n)\nonumber\\
&=&\frac{2}{\beta U}\sum_n\left [ -1+(i\omega_n+\mu-\Sigma_n)
G_n\right ]\Sigma_n.
\label{eq: 2corr_fun6}
\end{eqnarray}
This result is quite similar to the previous result for the other correlation
function, except now we have an extra weighting factor of $2\epsilon({\bf k})$
in the summation over momentum.

\section{The Falicov-Kimball model in infinite dimensions}

In this Section, we examine the time-dependence of the
local Green's function for the Falicov-Kimball model on an infinite-dimensional
hypercubic lattice.  We consider the case of half-filling with the system being 
coupled to an external homogeneous time-dependent electric field. The time
dependence is taken to be particularly simple, at $t=t_0$ a constant field
is instantly turned on.
The formalism involves generalizing the DMFT to the nonequilibrium case.
The way to do this is based on a Kadanoff-Baym approach in real time, where
continuous matrix operators are discretized along the Kadanoff-Baym contour,
and operator manipulations are carried out on the discretized matrices using
standard linear-algebra approaches.  A short description of this technique,
including a benchmark against the well-known equilibrium solutions has
already appeared\cite{Nashville}; further details of this approach will
appear elsewhere.
To test our formulas for the first few moments, we compare numerically calculated
local moments to the exact moments in
Eqs.~(\ref{mu0Rdensityhalffilling}--\ref{mu2Rdensityhalffilling}) and 
(\ref{mu0<density}--\ref{mu2<density}).

The action for the Falicov-Kimball model is quadratic in the conduction
electrons.  Hence the Feynman path integral over the Kadanoff-Baym contour
can be expressed by a determinant of a continuous matrix operator whose
arguments are defined on the contour. Since the concentration of static 
particles on each site is conserved, the trace over the fermionic variables
can be straightforwardly taken. This is what allows the nonequilibrium DMFT
problem to be solved, but the technical details are complicated. The
first thing that needs to be noted is that the self-energy remains local
even in the presence of a field.  This follows by applying Langreth's
rules\cite{langreth} 
to the perturbation theory, which state that every nonequilibrium
diagram can be related to an equilibrium diagram, but now one must perform
the analysis over the Kadanoff-Baym contour rather than  over the finite
imaginary time interval.  Since the perturbative analysis of the equilibrium
problem shows the self-energy to be local\cite{brandt_mielsch,metzner,Freericks} 
in equilibrium, it remains local in the nonequilibrium case as well. 

We couple the system in Eq.~(\ref{Hk}) to an
external electric field along the unit-cell diagonal direction in real space; this
yields
the following vector potential for the electric field\cite{Turkowski}:
\begin{equation}
{\bf A}(t)=A(t)(1,1,...,1).
\label{A}
\end{equation}
The bandstructure for noninteracting electrons coupled to the electric 
field in Eq.~(\ref{A}) has a simple form:
\begin{equation}
\epsilon \left({\bf k}-\frac{e{\bf A}(t)}{\hbar c}\right)
=\cos\left(\frac{eaA(t)}{\hbar c}\right)\varepsilon_{\bf k}
+\sin\left(\frac{eaA(t)}{\hbar c} \right){\bar \varepsilon}_{\bf k},
\label{energy}
\end{equation}
with the energy functions defined to be
\begin{equation}
\varepsilon_{\bf k}=-\frac{t^{*}}{\sqrt{d}}\sum_{j}\cos (a{\bf k}_{j})
\label{eps}
\end{equation}
and
\begin{equation}
{\bar \varepsilon}_{\bf k}=-\frac{t^{*}}{\sqrt{d}}\sum_{j}\sin (a{\bf k}_{j}),
\label{bareps}
\end{equation}
and $t^{*}$ is the renormalized
hopping parameter\cite{metzner_vollhardt} in the limit of $d\rightarrow\infty$;
we take $t=t^{*}/2\sqrt{d}$.

Because the Green's functions now depend on two energies, the summation over
the infinite-dimensional Brillouin zone can be replaced by a double integral
over a joint density of states for the two energies.
This joint density of states (DOS) is\cite{SchmidtMonien}
\begin{eqnarray}
\rho_{2}(\varepsilon , {\bar \varepsilon})
=\frac{1}{\pi t^{*2}a^{d}}\exp\left[ 
-\frac{\varepsilon^{2}}{t^{*2}}-\frac{{\bar \varepsilon}^{2}}{t^{*2}}
\right] .
\label{rho2}
\end{eqnarray}
The numerical integration over the joint DOS is performed by an averaged
Gaussian integration with 54 and 55 points for each energy axis:
\begin{equation}
\int_{-\infty}^{\infty}d\varepsilon \exp (-\varepsilon^{2})F(\varepsilon )
\simeq \sum_{i=1}^{N}w_{i}F(\varepsilon_{i}),
\label{Gaussian}
\end{equation}
where $w_{i}$ are Gaussian weights which correspond to the $N$ energy
points $\varepsilon_{i}$. Since the Green's functions
often depend on the energy as $\exp (ic\varepsilon)$,
the Gaussian quadrature rule in Eq.~(\ref{Gaussian})
fails to give correct results when $c$ is on the order of (or larger than)
the inverse
of the grid spacing of the energy points near $\varepsilon =0$.
In this case, the sum over discrete points
contains a systematic contribution of terms
which do not cancel each other, and leads to an overestimated value
to the integral. One way to efficiently correct this is to average 
two Gaussian summations with numbers of Gaussian points
equal to $N$ and $N+1$ because the Gaussian points interleave each other, and
act like a step size about half as big as either sum alone, and they give 
somewhat better accuracy than performing the integral with $2N+1$ points,
because a subset of those points are at such large absolute value that
the Gaussian weight is small enough that it can be safely neglected.

\begin{figure}[h]
\centering{
\includegraphics[width=1.5in,angle=270]{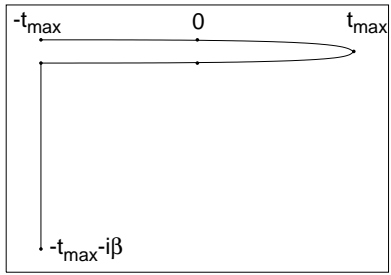}}
\caption{
Kadanoff-Baym contour for the two-time Green's functions
in the nonequilibrium case. We take the contour to run from $-t_{max}$ to
$t_{max}$ and back, and then extends downward parallel to the imaginary
axis for a distance of $\beta$.  The field is usually turned on at $t=0$;
{\it i.e.}, the vector potential is nonzero only for positive times.
}  
\label{fig: contour}
\end{figure}

In order to calculate the nonequilibrium local Green's functions, one
needs to self-consistently solve a system of equations\cite{Nashville}
 which connects these functions with the corresponding 
local self-energy $\Sigma (t_{1},t_{2})$ and an effective dynamical
mean-field $\lambda (t_{1},t_{2})$;
these equations are similar in form to the equilibrium 
case,\cite{Georges,Freericks}~but now all the functions 
are discrete matrices of two time variables defined
on the Kadanoff-Baym contour (see Fig.~\ref{fig: contour}).
Details of the algorithm and the nonequilibrium DMFT
equations will be given elsewhere\cite{unpublished}.

\begin{figure}[h]
\centering{
\includegraphics[width=3.0in,clip=]{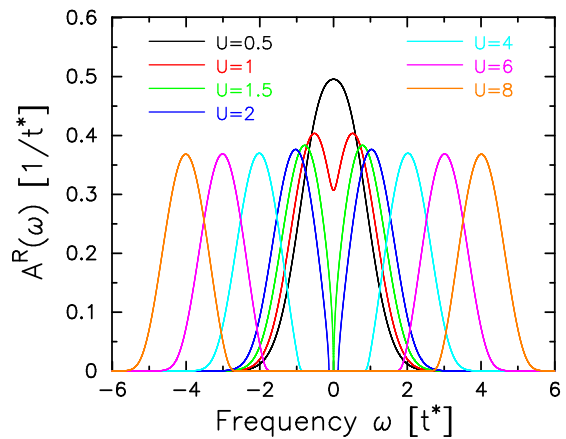}}
\caption{DOS for the equilibrium retarded Green's function
for different values of $U$.  The DOS is independent of temperature.
}  
\label{fig: ret_eq}
\end{figure}

\begin{figure}[h]
\centering{
\includegraphics[width=3.0in,clip=]{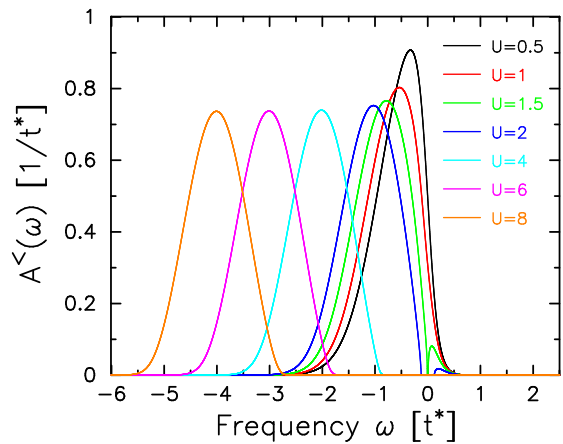}}
\caption{Local lesser Green's function in equilibrium
at $\beta =10$ and for different values of $U$. 
}  
\label{fig: less_eq}
\end{figure}

First, we present results for the local moments in 
equilibrium, when the system of the DMFT equations
is solved in the frequency representation using the Brandt-Mielsch 
approach\cite{brandt_mielsch}
(for details, see Ref.~\onlinecite{Freericks}).
Plots of the local retarded and lesser DOS for different values
of $U$ are shown in Figs.~\ref{fig: ret_eq} and \ref{fig: less_eq}.
The metal-insulator transition occurs at $U= \sqrt{2}$;
the insulator has anomalous properties because there is no real gap---instead
the DOS is exponentially small in a gap region around the chemical potential,
and vanishes only at the chemical potential.

The moments for the retarded and lesser Green's functions are calculated
by directly integrating the Green's functions multiplied by the
corresponding power of frequency; we use a step size of 
$\Delta\omega =0.001$ and a rectangular quadrature rule.  The results
for half-filling are presented in Table~\ref{Tab1}.
One can immediately see that
the numerical results for the first moment of the lesser
Green's function
are in an excellent agreement with the exact expressions for the moment (the 
zeroth and second moments agreed exactly with their exact expressions).
We also calculated the retarded moments and they agreed exactly with
the exact values, so we don't summarize them in a table.
Note that there appears to be a relation between the second moments
of the retarded and lesser Green's functions at half filling.  This
relation ceases to hold off of half filling.  For example,
in the case with $U=2$, $w_1=0.25$, and $\rho_e=0.75$ we find the following 
moments (exact results in parentheses): 
$\tilde \mu_1^R=-0.872072144$ $(-0.872071963)$;
$\tilde \mu_2^R=2.01050948$ $(2.01050951)$;
$\tilde \mu_1^<=-2.149496$ $(-2.149462)$;
and $\tilde \mu_2^<=3.623019$ $(3.622970)$. These
results are more indicative of the general case (where 
$\tilde\mu_2^R\ne\tilde\mu_2^<$).  Note that one needs
to use many Matsubara frequencies in the summations to get good convergence
for the average kinetic energy and for the second correlation function
(we used 50,000 in this calculation with $\beta=10$).
The majority of our numerical error comes from the difficulty in exactly 
calculating those results; indeed, the exact result, calculated from the
operator averages on the Matsubara frequency axis is probably less accurate
than the direct integration of the moment on the real axis. One can improve 
this situation somewhat by working
on the real axis to calculate the different operator averages, but we wanted
to indicate the accuracy under the most challenging circumstances. We feel
our final results are quite satisfactory, and indicate these sum rules do
hold.

\begin{table*}
\caption{First spectral moment for the lesser Green's function in  equilibrium 
with $\beta =10$ and different values of $U$. The zeroth moment is
accurate to more than eight digits, and is not included. Similarly, we
find the second moment is equal to $0.5+U^2/4$ to high accuracy, and
is not included. Note that the first moment continuously evolves
from the value $-1/\sqrt{\pi}$ for $U=0$ and $\beta=\infty$ to approximately
$-U/2$ as $U$ increases. The approach to $-U/2$ is expected due to the
formation of upper and lower Hubbard bands separated by $U$.}
\label{Tab1}
\begin{ruledtabular}
\begin{tabular}{ccccccc}
moment                        &  $U=0.5$     & $U=1.0$      &  $U=1.5$       & $U=2.0$      & $U=4.0$ & $U=6.0$\\
${\tilde \mu}_{1}^{<}$        & -0.591699    & -0.717901    & -0.902869      & -1.119047    & -2.062036 & -3.041526\\
${\tilde \mu}_{1}^{<}(exact)$        &-0.591687     &-0.717886     &-0.902848       &-1.119017     &-2.061945 & -3.041333\\
\end{tabular}
\end{ruledtabular}
\end{table*}

\begin{figure}[h]
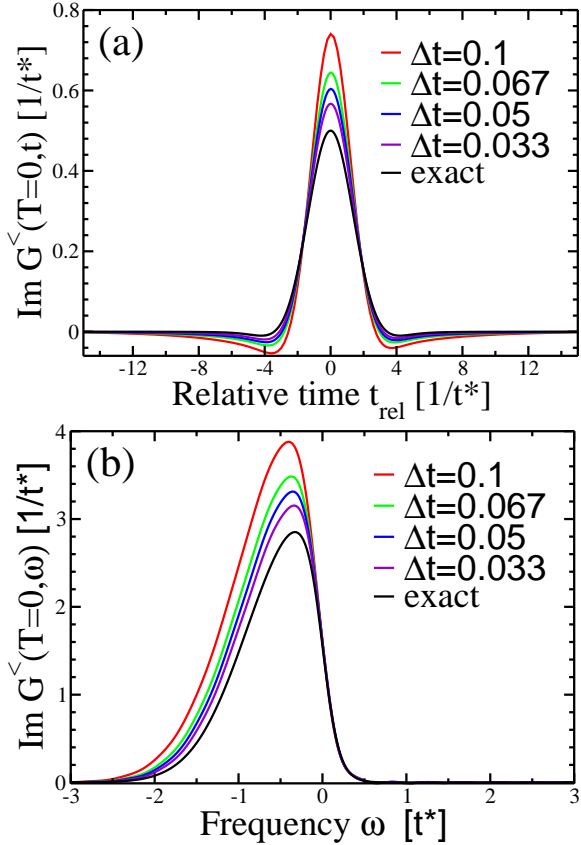

\centering{
\includegraphics[width=3.0in,clip=]{Fig4a}}
\centering{
\includegraphics[width=3.0in,clip=]{Fig4b}}
\caption{(Panel a) 
Imaginary part of the lesser Green's function as a function of the 
relative time coordinate for different discretizations of the time contour.
The model parameters are $U=0.5$, $\beta =10$ and $E=0$.
The average time $T$ is set equal to zero.
The parameters for the Kadanoff-Baym time-contour discretization are:
$t_{max}=15$ and $\Delta \tau=0.1$ ({\it i.e.}, 100 points taken along the
imaginary axis); the discretization along the real time
axis is given by $\Delta t$ as shown in the figure. Note how the results
systematically approach the exact result as the discretization goes to
zero. (Panel b) 
Imaginary part of the lesser Green's function as a function of 
frequency for different discretizations of the time contour.
}  
\label{fig: eq_lesser_0.5}
\end{figure}

We now focus on numerical results for the nonequilibrium code.  Our first
benchmark is to calculate equilibrium results with that code and compare
with exact results available for the equilibrium case.
Such calculations\cite{Nashville} show good convergence and
precision when $U$ lies below the critical $U$ for the metal-insulator 
transition ($U<\sqrt{2}$).
Here we demonstrate this by showing how the relative
time dependence of the imaginary part of the lesser Green's function
at $U=0.5$ and $\beta =10$ changes when one decreases the time step 
$\Delta t$ on the real part of the Kadanoff-Baym contour.
As follows from Fig.~\ref{fig: eq_lesser_0.5}, the solution becomes more 
accurate 
as $\Delta t$ decreases but the accuracy is reduced as the temperature is
lowered, as can be seen by comparing to the $\beta=1$ results in
Ref.~\onlinecite{Nashville}. For this calculation, we determine the moments by
taking the derivatives of the Green's functions in the time representation.
The values for the spectral moments
at $U=0.5$, $\beta =10$ and different values of $\Delta t$ 
are presented in Tables \ref{Tab3} and \ref{Tab4}.
The results for the moments improve as $\Delta t$ decreases.  

\begin{table}
\caption{Spectral moments for the retarded Green's function in the case
of zero electric field at $U=0.5$, 
$\beta =10$ and calculated with the nonequilibrium real-time formalism
with different values for the time step $\Delta t$.
The other parameters are $t_{max}=15$, $N=54,~55$, $\Delta \tau =0.1$
\label{Tab3}}
\begin{ruledtabular}
\begin{tabular}{cccccc}
moment                 & $\Delta t=0.1$ &  $\Delta t=0.067$ &  $\Delta t=0.05$ & $\Delta t=0.033$ & exact \\
${\tilde \mu}_{0}^{R}$ & 1.580785       &  1.331640         &  1.232022        & 1.144811         & 1 \\
${\tilde \mu}_{1}^{R}$ & 0.174040       &  0.082610         &  0.052785        & 0.030002         & 0\\
${\tilde \mu}_{2}^{R}$ & 1.324976       &  0.979230         &  0.848047        & 0.737020         & 0.5625\\
\end{tabular}
\end{ruledtabular}
\end{table}

\begin{table}
\caption{The same as in Table III but for the case of the lesser Green's function.
\label{Tab4}}
\begin{ruledtabular}
\begin{tabular}{cccccc}
moment                 & $\Delta t=0.1$ & $\Delta t=0.067$ & $\Delta t=0.05$ & $\Delta t=0.033$ &  exact \\
${\tilde \mu}_{0}^{<}$ & 1.480893       & 1.289036         & 1.207662        & 1.133850         &  1\\
${\tilde \mu}_{1}^{<}$ & -1.036753      & -0.850675        & -0.774525       & -0.706893        & -0.591687\\
${\tilde \mu}_{2}^{<}$ & 1.108705       & 0.870853         & 0.777152        & 0.695791         &  0.5625\\
\end{tabular}
\end{ruledtabular}
\end{table}

\begin{figure}[h]
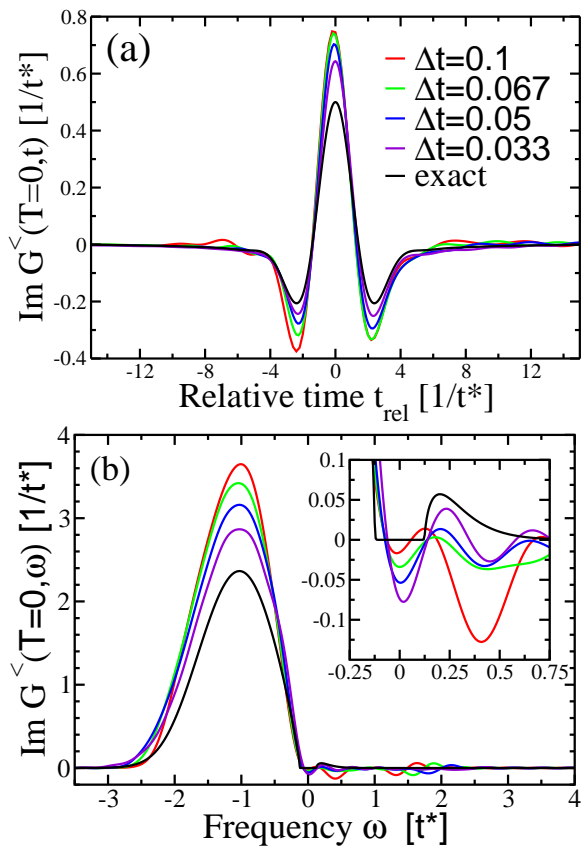

\centering{
\includegraphics[width=3.0in,clip=]{Fig5a}}
\centering{
\includegraphics[width=3.0in,clip=]{Fig5b}}
\caption{(Panel a)
Imaginary part of the lesser Green's function as a function of the
relative time coordinate for different discretizations of the time contour.
The model parameters are $U=2$, $\beta =10$ and $E=0$.
The average time $T$ is set equal to zero.
The parameters for the Kadanoff-Baym time-contour discretization are:
$t_{max}=15$ and $\Delta \tau=0.1$ ({\it i.e.}, 100 points taken along the
imaginary axis); the discretization along the real time
axis is given by $\Delta t$ as shown in the figure. Note how the results
systematically approach the exact result as the discretization goes to
zero. (Panel b)
Imaginary part of the lesser Green's function as a function of
frequency for different discretizations of the time contour; in the inset,
the region around $\omega=0$ is blown up to show the gap development as a
function of the discretization. Note how the
gap region converges very slowly---instead we see the DOS go negative in the
gap region.  Properly determining that structure in the frequency domain
requires the Green's function over an extended time domain, which is not
numerically feasible.
\label{fig: eq_lesser_2}}
\end{figure}

In the insulating phase, the calculations are less accurate 
(see Fig.~\ref{fig: eq_lesser_2}). The self-energy
develops a pole in the frequency representation, which gives the imaginary
part a delta function at the chemical potential.  Hence, we expect a constant
background value for the self-energy in the time representation.  This cannot
be properly represented in a numerical calculation that has a finite cutoff in
the time domain, which makes the real-time formalism much more challenging in 
the insulating phase.  The problems appear to be somewhat less pronounced
for the Green's functions, but the convergence is slower than in the metallic
phase, and the gap region has unphysical behavior [the DOS goes negative in
the gap region as shown in the inset to panel (b)]. One cannot get rid of these
oscillations without having a time-domain cutoff that extends to infinity, but
by comparing with the exact results, we find that a time-domain cutoff of about
$t_{rel}\approx 200$ provides quite reasonable results for the calculations;
in our results with the nonequilibrium code, we are limited to time-domain cutoffs
of closer to $15-30$, which explains the poor agreement for the gap region.
Fortunately, these numerical problems appear to reduce when
an external field is turned on, and we are in the nonequilibrium case.

Note that we show equilibrium
results only for the average time $T=0$.  In equilibrium, the results should be
independent of $T$, but we find that we have a modest $T$ dependence due
to discretization error.  The results at $T=0$ turn out to be the least
accurate, and the accuracy of the results improves as the discretization
size is made smaller.  In general, we find the $T$ dependence of the Green's
functions to vary (pointwise) by no more than 40\% for $\Delta t=0.1$ and to be
reduced to a 10\% variation when $\Delta t=0.033$ (for $U=0.5$). The variation
is about three times larger for $U=2$.  We find less variation in the
nonequilibrium calculations, which appear to be better suited to the
real-time formalism [most likely because the Green's functions don't behave
like complex exponentials $\exp(i\epsilon t)$ as the equilibrium functions do;
such functions can be particularly difficult to deal with in our real-time
numerical calculations].

\begin{figure}[h]
\centering{
\includegraphics[width=3.0in,clip=]{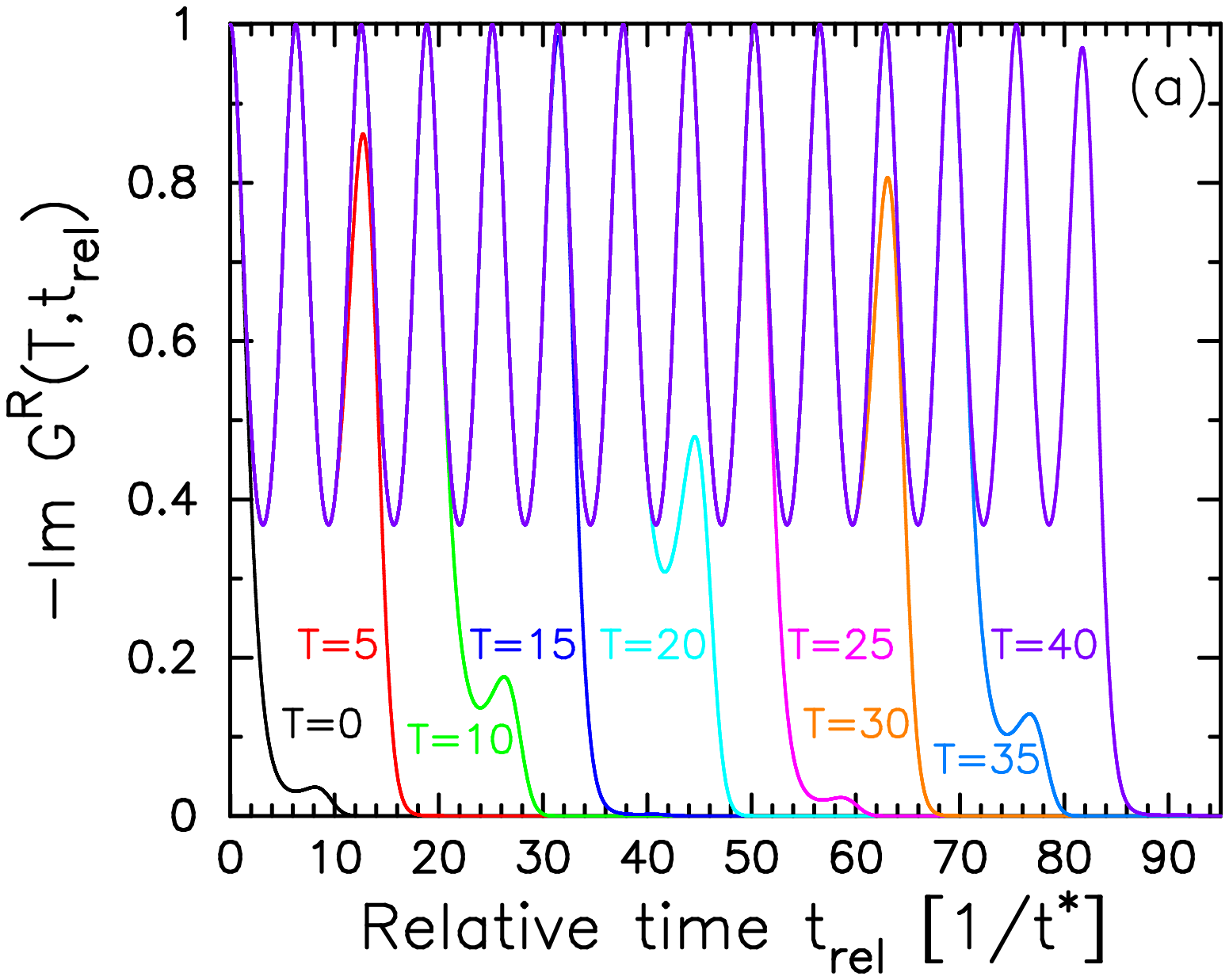}
\includegraphics[width=3.0in,clip=]{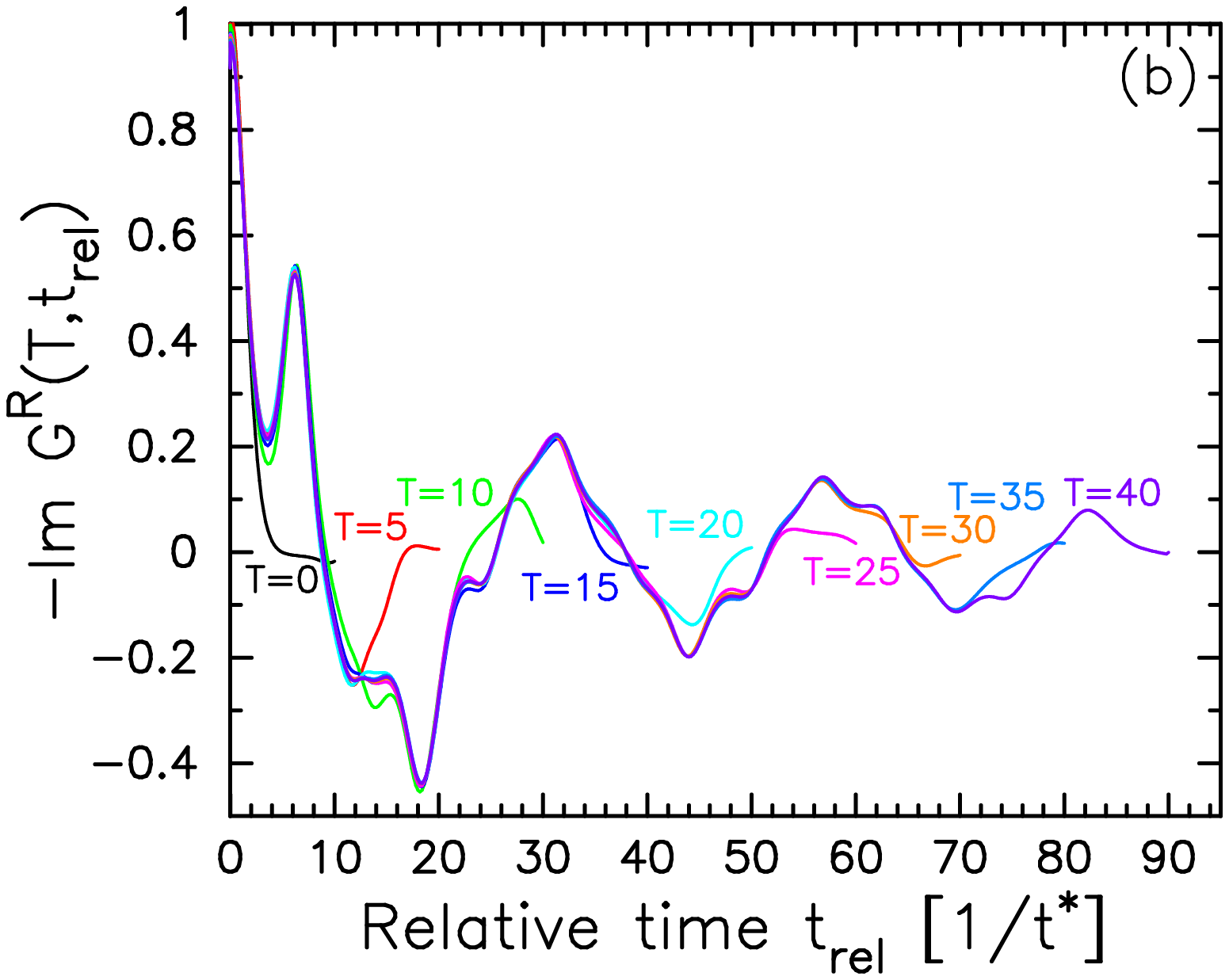}}
\caption{Imaginary part of the retarded Green's function as a function of the 
relative time coordinate at different values of the average time (starting at
$T=0$ and running to $T=40$ in steps of 5).
The field is switched on at the time $T=0$ and we take $\Delta t=0.1$ along the
Kadanoff-Baym contour.  The model parameters are $\beta =10$ and $E=1$.
Panel (a) shows the noninteracting result $U=0$, while panel (b) shows the 
interacting result $U=0.5$ (note that the curves extend as far out in 
$t_{rel}$ as we have data; the cutoff in $t_{rel}$ comes from the finite
time domains of our calculations).
Note how the results appear to retrace themselves for different average times
until the relative time becomes larger than approximately $2T$, where the
Green's function decays.  For the noninteracting case, the pattern obviously
becomes periodic in the Bloch period as $T\rightarrow\infty$, which leads to delta
functions in the Fourier transform (with respect to $t_{rel}$), whereas the 
interacting case may or may not
be approaching a periodic form for large relative time; the data does not
extend far enough out to be able to make a conclusion about the
asymptotic form.
\label{fig: noneq_u=0.5_e=1}}
\end{figure}

As a nonequilibrium problem, we consider the case
of the interacting Falicov-Kimball model in the metallic phase 
at $U=0$ and $U=0.5$ with $\beta =10$ 
and a constant electric field $E=1$ is turned on at some moment
of time.  The Green's functions for noninteracting
electrons\cite{Turkowski} [see panel (a) of Fig.~\ref{fig: noneq_u=0.5_e=1}]
are oscillatory functions of the relative time coordinate
in presence of the external field. The average time dependence is weak for 
relative times up to about $2T$, after which, the Green's function decays
as a function of $t_{rel}$. Notably, the results for different average times
lie on top of each other until the relative time becomes larger than $2T$.  This
implies that there is little or no average time dependence to the retarded Green's
function as $T\rightarrow\infty$, and the retarded Green's function becomes
a periodic function of $t_{rel}$ (with the Bloch period); this latter
result implies that the Fourier transform will consist of evenly spaced
delta functions, which is the familiar Wannier-Stark ladder. When interactions are
turned on, the situation changes [see panel (b) of 
Fig.~\ref{fig: noneq_u=0.5_e=1}].  We still see little average time dependence
for relative times smaller than $2T$, but the Green's function does not
appear periodic in $t_{rel}$ for small times.  The critical question to ask is,
what happens as $T\rightarrow\infty$?  If the Green's function becomes periodic
in $t_{rel}$, then there will be delta functions surviving in the
Fourier transform, but if it continues to decay, there will not. Our data
does not extend far enough out in time for us to be able to resolve this
issue.

\begin{figure}[h]
\centering{
\includegraphics[width=3.0in,clip=]{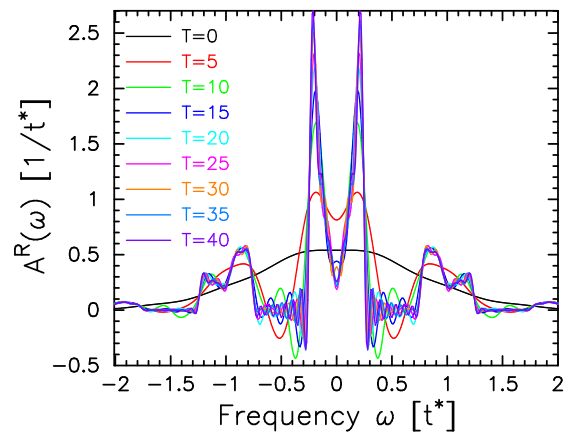}}
\caption{Density of states for different average times (running from
$T=0$ to $T=40$ in steps of 5) for the interacting system, with $U=0.5$ and
$E=1$.  These results correspond to the Fourier transform of the data in
Fig.~\ref{fig: noneq_u=0.5_e=1}.
Note how the DOS seems to be approaching a limiting form even for
this small a value of the average time.  We cannot tell, however, whether there
might be some low-weight delta functions appearing somewhere in the spectrum.
In any case, it does appear that there are no sharp structures near the Bloch
frequencies, which are at integer frequencies for $E=1$.
\label{fig: dos_u=0.5_e=1}}
\end{figure}

It is also interesting to examine the density of states in frequency space.
The $U=0$ case has been studied exhaustively~\cite{Turkowski},
so we don't repeat it here.  The interacting case is plotted in
Fig.~\ref{fig: dos_u=0.5_e=1}.  In constructing this plot, we set
the real part of the retarded Green's function to zero before performing
the Fourier transform, since the exact result must vanish by particle-hole
symmetry, and our numerical calculations have a small nonzero real part.
Note how the DOS rapidly readjusts itself into a nonequilibrium form, and how
the steady state appears to be approximately reached.  The only subtle issue is 
the one discussed above, of whether there will be delta function peaks emerging
in the DOS as the average time gets larger.  What is clear, is that if such peaks
do form, they are quite unlikely to appear at multiples of
the Bloch frequencies, because
we see no sharp peaks forming near integer frequencies here (the Bloch frequency
is equal to 1 for $E=1$); indeed, the DOS seems to be suppressed at integer
frequencies.

\begin{figure}[h]
\centering{
\includegraphics[width=3.0in,clip=]{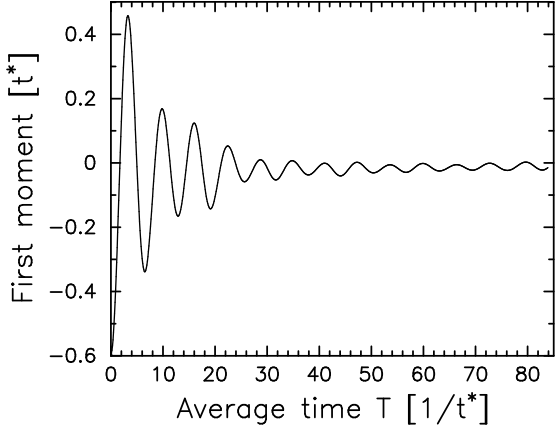}}
\caption{ First moment of the local lesser function $\tilde \mu^<_1$ plotted
as a function of average time T for $U=0.5$ and $E=1$.  The field is turned on
at $T=0$. The step size is
$\Delta t=0.1$, and the moment is calculated from the numerical derivative
of the data.
\label{fig: first_moment}}
\end{figure}

The lesser Green's functions are remarkably similar to the retarded Green's 
functions, except they are nonzero for all $t_{rel}$, not just for positive 
values.  There also is limited average time dependence, except in the region
of small $t_{rel}$, where the first derivative of the Green's function
does vary with average time.  It is this variation that leads to oscillations
in the current, and is critical for understanding the behavior of these
systems.  Unfortunately, it is not simple to directly evaluate such derivatives
accurately when we calculate them numerically, because our time step is
rather large.  We examine them in this numerical fashion, 
with results plotted in Fig.~\ref{fig: first_moment} for the case $\Delta t=0.1$.
Note how the first moment oscillates, then decays, and finally seems to reach
a steady oscillatory state as the average time increases.
We cannot tell whether the moment becomes a constant at large average
times or continues to oscillate from the data that we have, although it appears
to be approaching an oscillatory steady state. Note further that because $U$ is 
not too large here, we anticipate a correlation between these oscillations in the 
first moment and oscillations of the current.  Indeed, if one calculates the
current, one finds that it also appears to approach a steady oscillating state for
large average time; details of these results will be presented elsewhere.

Despite a significant change in the Green's functions after switching on
an electric field (compare the $T=0$ results, which are close to the equilibrium
results to the larger $T$ values in Figs.~\ref{fig: noneq_u=0.5_e=1}
and \ref{fig: dos_u=0.5_e=1}), 
the spectral moments for these
functions do not change much (Tables \ref{Tab5} and \ref{Tab6}).
The spectral moments are connected with the relative time derivatives
of the Green's functions [Eq.~(\ref{munR<5})]; the fact that some moments
do not change in the presence of an electric field suggests that      
the those derivatives are independent of the electric field. Indeed, 
we calculate the moments in these examples from the derivatives of the 
Green's functions at $t=0$ because we often do not have data out to a
large enough relative time to perform the Fourier transform, and evaluate
the moment in the conventional way.
Note that, the second moment, which is actually equal
to the curvature of $G^{R}$ and $G^{<}$, is independent of average time
for the retarded function, and does not appear to have average time
dependence for the lesser function. The first moment of the lesser function
does depend on average time (see Fig.~\ref{fig: first_moment}).
Finally, it is interesting to observe that the values for the moments
are much closer to the exact results for a similar discretization size
than what we found in the equilibrium case.  This is why we believe that the
real-time numerical algorithm converges better for the nonequilibrium case than
for the equilibrium case.

\begin{table}
\caption{Spectral moments for the retarded Green's function in the case
when the constant external electric field $E=1$ is switched on at $T=0$.
The parameters are $U=0.5$, $\beta =10$,
$\Delta t=0.05$, $t_{max}=15$, $N=54,~55$, and $\Delta \tau =0.05$.
These moments should all be independent of time, and they appear to be
within the numerical error.
}
\label{Tab5}
\begin{ruledtabular}
\begin{tabular}{ccccccc}
moment                 &  T=0  & T=5 & T=10 & T=15 & T=20 & exact \\
${\tilde \mu}_{0}^{R}$ & 1.0025  & 1.0088 & 0.9985 & 0.9951 & 0.9967  & 1     \\
${\tilde \mu}_{1}^{R}$ & 0.00665 & -0.00054 & 0.00005 & 0.00054 & 0.00003 & 0     \\
${\tilde \mu}_{2}^{R}$ & 0.56155 & 0.56198 & 0.55184 & 0.55030 & 0.55112 & 0.5625\\
\end{tabular}
\end{ruledtabular}
\end{table}

\begin{table}
\caption{The same as in Table IV for the case of the lesser Green's function.
The first moment appears to change with average time in the nonequilibrium
case.  The equilibrium value is $-0.5917$, which agrees well with our
result before the field is turned on. The second moment appears to be 
independent of average time, even in a field, but we have no proof of this 
fact.}
\label{Tab6}
\begin{ruledtabular}
\begin{tabular}{ccccccc}
moment                 &  T=0  & T=5 & T=10 & T=15 & T=20 & exact \\
${\tilde \mu}_{0}^{<}$ & 1.0025  & 1.0098 & 0.9997 & 0.9960 & 0.9975  & 1     \\
${\tilde \mu}_{1}^{<}$ &-0.5878  & 0.0042 &-0.1618 &-0.0454 & 0.0933 & ?
  \\
${\tilde \mu}_{2}^{<}$ & 0.5520 & 0.5666 & 0.5554 & 0.5572 & 0.5588 & ?\\
\end{tabular}

\end{ruledtabular}
\end{table}

To conclude our numerical analysis, we consider
the spectral moments of the lesser Green's function 
in the case where it is approximated by the
generalized Kadanoff-Baym (GKB) ansatz\cite{Lipavsky}. The idea of the GKB is
to represent the lesser Green's function in terms of the distribution function
(determined by the $t=0$ limit of the lesser Green's function) and
the full two-time retarded and advanced Green's functions:
\begin{eqnarray}
\hat G_{{\bf k}}^{<}(t_{1},t_{2})&=&-i\left[
G_{{\bf k}}^{R}(t_{1},t_{2})G_{{\bf k}}^{<}(t_{2},t_{2})\right.\nonumber\\
&-&\left. G_{{\bf k}}^{<}(t_{1},t_{1})G_{{\bf k}}^{A}(t_{1},t_{2})
\right] .
\label{KadanoffBaym}
\end{eqnarray}
The expression for the spectral moments of the lesser Green's function,
approximated by Eq.~(\ref{KadanoffBaym}),
can be found by taking relative time derivatives
of $\hat G_{{\bf k}}^{<}(T,t)$ as shown in Eq.~(\ref{munR<5}). 
Similar to the derivation performed in Section II, one finds (the hat denotes 
the GKB approximation)
\begin{eqnarray}
\hat \mu_{0}^{<}({\bf k},T)&=&2n_{c}({\bf k},T),
\label{mu0<GKB}\\
\hat \mu_{1}^{<}({\bf k},T)&=&n_{c}({\bf k},T)
\left(
\left\langle\left\{ [c_{{\bf k}}(T),H],c_{{\bf k}}^{\dagger}(T)\right\}
\right\rangle\right.
\nonumber\\
&-&\left.
\left\langle\left\{ c_{{\bf k}}(T),[c_{{\bf k}}^{\dagger}(T),H]
\right\}\right\rangle\right) 
\nonumber \\
&+&{\rm Im}\frac{\partial n_{c}({\bf k},T)}{\partial T},
\label{mu1<GKB}\\
\hat \mu_{2}^{<}({\bf k},T)
&=&\frac{1}{2}n_{c}({\bf k},T)
\left(
\left\langle\left\{ [[c_{{\bf k}}(T),H],H],c_{{\bf k}}^{\dagger}(T)\right\}
\right\rangle
\right.
\nonumber \\
&-&2\left. \left\langle\left\{ [c_{{\bf k}}(T),H]
,[c_{{\bf k}}^{\dagger}(T),H]\right\}
\right\rangle\right. \nonumber\\
&+&\left. \left\langle\left\{ c_{{\bf k}}(T),[[c_{{\bf k}}^{\dagger}(T),H],H]
\right\}\right\rangle\right) \nonumber\\
&-&\frac{1}{2}{\rm Re}
\frac{\partial^{2}n_{c}({\bf k},T)}{\partial T^{2}}.
\label{mu2<GKB}
\end{eqnarray}
Comparison of the expressions in Eqs.~(\ref{mu0<GKB})--(\ref{mu2<GKB}) 
for the GKB moments
with the corresponding exact expressions in Eqs.~(\ref{mu0R}), (\ref{mu1Rder3})
and (\ref{mu2Rderiv2}) for the moments of $G_{{\bf k}}^{R}$
allows us to connect
the lesser GKB spectral moments with the exact retarded spectral moments:
\begin{equation}
\hat \mu_{n}^{<}({\bf k},T)=2n_{c}({\bf k},T)\mu_{n}^{R}({\bf k},T)
-\delta_{n,2}\frac{1}{2}
\frac{\partial^{2}n_{c}({\bf k},T)}{\partial T^{2}}, 
\label{GKBretarded}
\end{equation}
for $n=0$, 1, and 2.
Since the time derivatives of the momentum distribution function
are real valued, the last term in Eq.~(\ref{GKBretarded}) 
is nonzero only for $n=2$.
Thus, the lesser GKB spectral moments (with $n=0$, 1, and 2)
are equal to the corresponding retarded spectral moments
multiplied by $2n_{c}({\bf k},T)$ plus a term
involving the second time derivative of the momentum distribution function 
for the case of $n=2$.

Comparison of Eq.~(\ref{GKBretarded}) and the
exact expressions for the retarded spectral moments [in
Eqs.~(\ref{mu0R}), (\ref{mu1R1}) and (\ref{mu2R})] with the exact
expressions for the lesser moments [in Eqs.~(\ref{mu0<}), (\ref{mu1<})
and (\ref{mu2<})] allows us to conclude that the lesser spectral
moments for the GKB approximated Green's function
correspond to an approximation of the exact spectral moments by evaluating
the operator averages with a mean-field approximation
(the second moment contains an additional term with
the second time derivative of the momentum distribution function). This indicates
that the GKB approximation will fail as the correlations increase, because
it produces the wrong moments to the spectral functions. Note that this does not
say the GKB approach is a mean field theory approach, it is not.

\section{Conclusions}

In this paper, we derived a sequence of spectral moment sum rules for the
retarded and lesser Green's functions of the Falicov-Kimball and Hubbard models.
Our analysis holds in equilibrium and in the nonequilibrium case of a spatially
uniform, but time-dependent external electric field being applied to the system.
Our results are interesting, because they show there is no average time dependence
nor electric field dependence to the first three moments of the local 
retarded Green's
functions.  This implies that the value, slope and curvature of the local retarded
Green's
functions, as functions of the relative time, do not depend on the average
time or the field.  Such a result extends the well-known result that the
total spectral weight (zeroth moment)
of the Green's function is independent of the field.
It also implies that one will only see deviations of Green's functions from
the equilibrium results when the relative time becomes large. Such an observation
is quite useful for quantifying the accuracy of nonequilibrium calculations.
We showed some numerical results illustrating this effect for nonequilibrium
DMFT calculations in the Falicov-Kimball model.
The case for the lesser Green's function is more complicated, and there is both
average time dependence and field dependence apparent in those moments, although
it appears that the second moment at half filling may not depend on average 
time.

We also examined a common approximation employed in nonequilibrium calculations,
the so-called generalized Kadanoff-Baym ansatz.  We find the moments in that
case are similar in form to the exact moments, except they involve a
mean-field-theory decoupling of correlation function expectation values
when one evaluates the operator averages that yield the sum-rule values.
This implies that the GKB approximation must fail as the correlations increase,
and such a mean-field decoupling becomes inaccurate, because it will have
the wrong spectral moments.

We hope that use of these spectral moments will become common in nonequilibrium
calculations in order to quantify the errors of the calculations for small
times. We believe they can be quite valuable in checking the fidelity of
numerical calculations and of different kinds of approximate solutions.

\section*{Acknowledgments}

We would like to acknowledge support by the National Science Foundation under 
grant number DMR-0210717 and by the Office of Naval Research under grant
number N00014-05-1-0078.

\appendix

\section{Derivation of the second spectral moment for the retarded
Green's function}

We show the explicit steps needed to derive Eq.~(\ref{mu2Rderiv2})
by using Eqs.~(\ref{munR<}) and (\ref{GR})
for the second spectral moment of the retarded Green's function.
In this case, the second spectral moment is equal to
\begin{widetext}
\begin{equation}
\mu_{2}^{R}({\bf k},T)=
\left( -\frac{1}{\pi}\right)
\int_{-\infty}^{\infty}d\omega
{\rm Im}\int_{-\infty}^{\infty} dt e^{i\omega t}
i\frac{\partial^{2}}{\partial t^{2}}
\left[
\theta (t)
\left\langle\left\{ c_{{\bf k}}\left(T+\frac{t}{2}\right) 
   ,c_{{\bf k}}^{\dagger}\left( T-\frac{t}{2}\right) \right\}\right\rangle 
\right] .
\label{A0}
\end{equation}
This expression is equivalent to
\begin{eqnarray}
\mu_{2}^{R}({\bf k},T)&=&-\frac{1}{\pi}\int_{-\infty}^{\infty}d\omega
{\rm Im}\int_{-\infty}^{\infty} dt 
e^{i\omega t}i\frac{\partial\delta (t)}{\partial t}
\left\langle\left\{ c_{{\bf k}}\left(T+\frac{t}{2}\right) 
   ,c_{{\bf k}}^{\dagger}\left( T-\frac{t}{2}\right) \right\}\right\rangle
\nonumber \\
&-&\frac{1}{\pi}\int_{-\infty}^{\infty}d\omega
{\rm Im}\int_{-\infty}^{\infty} dt e^{i\omega t}\delta (t)
\left[
\left\langle\left\{ i\frac{\partial}{\partial t}c_{{\bf k}}
\left(T+\frac{t}{2}\right) 
   ,c_{{\bf k}}^{\dagger}\left( T-\frac{t}{2}\right) \right\}\right\rangle
\right.
+\left.  \left\langle\left\{ c_{{\bf k}}\left(T+\frac{t}{2}\right)  
   ,i\frac{\partial}{\partial t}c_{{\bf k}}^{\dagger}
\left( T-\frac{t}{2}\right) 
\right\}\right\rangle
\right]
\nonumber \\
&+&\frac{1}{\pi}
\int_{-\infty}^{\infty}d\omega
{\rm Im}\int_{-\infty}^{\infty} dt e^{i\omega t}
i\theta (t)
\left[
\left\langle\left\{ 
i^{2}\frac{\partial^{2}}{\partial t^{2}}
c_{{\bf k}}\left(T+\frac{t}{2}\right) 
   ,c_{{\bf k}}^{\dagger}\left( T-\frac{t}{2}\right) 
\right\}\right\rangle
\right.
\nonumber \\
&+&\left.
2\left\langle\left\{i\frac{\partial}{\partial t} c_{{\bf k}}\left(T+\frac{t}{2}\right)  
   ,i\frac{\partial}{\partial t}
c_{{\bf k}}^{\dagger}\left( T-\frac{t}{2}\right) 
\right\}\right\rangle
+\left\langle\left\{ c_{{\bf k}}\left(T+\frac{t}{2}\right) 
    ,i^{2}\frac{\partial^{2}}{\partial t^{2}}
     c_{{\bf k}}^{\dagger}\left( T-\frac{t}{2}\right) \right\}\right\rangle 
\right] .
\label{A1}
\end{eqnarray}
\end{widetext}

The first two integrals in 
Eq.~(\ref{A1}) are equal to zero, that is,
\begin{eqnarray}
&-&\frac{1}{\pi}
\int_{-\infty}^{\infty}d\omega
{\rm Im}\int_{-\infty}^{\infty}dt e^{i\omega t}i
\frac{\partial\delta (t)}{\partial t}\nonumber\\
&\times&
\left\langle\left\{ c_{{\bf k}}\left(T+\frac{t}{2}\right) 
   ,c_{{\bf k}}^{\dagger}\left( T-\frac{t}{2}\right) \right\}\right\rangle
=0,
\label{A2}
\end{eqnarray}
and
\begin{eqnarray}
&-&\frac{1}{\pi}
\int_{-\infty}^{\infty}d\omega
{\rm Im}\int_{-\infty}^{\infty}dt e^{i\omega t}
\delta (t)\nonumber\\
&\times&\left[
\left\langle\left\{ i\frac{\partial}{\partial t}c_{{\bf k}}
\left(T+\frac{t}{2}\right) 
   ,c_{{\bf k}}^{\dagger}\left( T-\frac{t}{2}\right) \right\}\right\rangle
\right.
\nonumber \\
&+&\left.
\left\langle\left\{ c_{{\bf k}}\left(T+\frac{t}{2}\right)  
   ,i\frac{\partial}{\partial t}c_{{\bf k}}^{\dagger}
\left( T-\frac{t}{2}\right) 
\right\}\right\rangle
\right] =0.
\label{A3}
\end{eqnarray}
To prove Eq.~(\ref{A2}), 
we note that the delta-function derivative satisfies 
\begin{equation}
\int_{-\infty}^{\infty}dt
f(t)\frac{\partial\delta (t-a)}{\partial t}
=-f'(a).
\label{A5}
\end{equation}
Hence, we can transfer the time derivative of the delta function into a
time derivative of the other factors in the integrand.  This time derivative
has two terms: (i) the first term is the derivative of the exponential,
which introduces an additional factor of $i\omega$ (the operator average then
becomes trivial when we set $t=0$, since the anticommutator is equal to
1) and (ii) the second term
which involves derivatives of the creation and annihilation operators. 
Performing the integration over $t$ by using the delta function then yields
\begin{eqnarray}
&-&\frac{1}{\pi}
\int_{-\infty}^{\infty}d\omega
\Biggr ( {\rm Im} (\omega)
-{\rm Im}\left[
\left\langle\left\{ [c_{{\bf k}}(T),\mathcal{H(T)}]
   ,c_{{\bf k}}^{\dagger}(T) \right\}\right\rangle\right.
\nonumber \\
&~&-\left.
\left\langle\left\{ c_{{\bf k}}(T)
   ,[c_{{\bf k}}^{\dagger}(T),\mathcal{H(T)}]
\right\}\right\rangle
\right]\Biggr ), 
\end{eqnarray}
where we replaced derivatives with respect to time by commutators with the
Hamiltonian.
The first term has no imaginary part, so it vanishes, as do the second two
terms, since one can easily show the difference of the two operators is Hermitian,
and hence has a real expectation value.  This completes the proof of 
Eq.~(\ref{A2}).

To prove (\ref{A3}), we first perform the integration over $t$.
The result is equal to
\begin{eqnarray}
&-&\frac{1}{\pi}
\int_{-\infty}^{\infty}d\omega
{\rm Im} 
\frac{1}{2}\left[
\left\langle\left\{ i\frac{\partial }{\partial T}
c_{{\bf k}}(T),c_{{\bf k}}^{\dagger}( T) 
\right\}\right\rangle\right.\nonumber\\
&-&\left.\left\langle\left\{ c_{{\bf k}}(T),i\frac{\partial }{\partial T}
c_{{\bf k}}^{\dagger}( T) 
\right\}\right\rangle
\right]
\nonumber \\
&=&-\frac{1}{\pi}
\int_{-\infty}^{\infty}d\omega
{\rm Im} 
\frac{1}{2}
\left\langle\left[ 
\left\{ [c_{{\bf k}}(T),\mathcal{H}(T)] ,c_{{\bf k}}^{\dagger}( T)\right\} 
\right.\right.\nonumber\\
&-&\left.\left.
\left\{c_{{\bf k}}(T),[c_{{\bf k}}^{\dagger}(T),\mathcal{H}(T)]\right\}
\right]\right\rangle .
\label{A7}
\end{eqnarray}
The operator 
in the second line of Eq.~(\ref{A7}) is the same operator we saw above;
it is Hermitian so the imaginary part of the statistical average vanishes.
This proves (\ref{A3}). 

To complete the derivation of the second spectral moment, we note that only the 
last integral term in Eq.~(\ref{A1}) can be nonzero.
Substituting the operator time derivatives by their
commutators with the Hamiltonian finally
gives the expression in Eq.~(\ref{mu2Rderiv2})
of the second spectral moment for the retarded Green's function.

\end{document}